\documentstyle[seceq,supplement,epsfig]{ptptex}
\input{epsf.tex}

\title{
Monopole condensation and colour confinement
}

\author{
Adriano {\sc Di Giacomo}\footnote{
Talks delivered atthe 1997 Yukawa International seminar on
{\em NON PERTURBATIVE QCD. Structure of QCD vacuum.}
Kyoto Dec. 2-12-1997}
}
\inst{
Dipartimento di Fisica, Pisa University and INFN.
}

\recdate{
\today
}

\abst{
The evidence is reviewed for the mechanism of colour confinement in
QCD by dual superconductivity of the ground state, i.e. by condensation
of monopoles.
}

\begin{document}

\maketitle

\section{Introduction}
Most of the existing tests of Quantum Chromodynamics (QCD) as theory of strong
interactions come from short distance phenomena (Deep inelastic scattering,
$e^+\,e^-\to$~ hadrons, 
Jet physics,\ldots). Perturbation theory is supposed to
work in that regime because of asymptotic freedom.

On the other hand it is known that the renormalized perturbative expansion does
not converge, not even as an asymptotic series, due to bad infrared
behaviour\cite{1}. Fock vacuum is unstable: quarks and gluons, which are the
elementary excitations of perturbation theory, never appear as asymptotic
states.
This phenomenon is known as colour confinement.

A convincing experimental evidence for colour confinement is the upper limit
on the cosmic abundance of relic quarks, $n_q$:
\begin{equation}
\frac{n_q}{n_p} < 10^{-27}\label{eq:1.1}\end{equation}
$n_p$ is the abundance of nucleons.

Eq.(\ref{eq:1.1}) correspond to Millikan like analysis of $\sim 10^2$~gr of
matter. For non confined quarks the standard cosmological model predicts
\begin{equation}
\frac{n_q}{n_p} < 10^{-12}\label{eq:1.2}\end{equation}
A non perturbative formulation of the theory is needed, as well as a
theoretical understanding of why perturbation theory works at all at short
distances.

Lattice formulation provides that formulation.

The Feynman path integral defining the theory is regularized by discretizing
space time, and computed numerically by Montecarlo techniques. Of course
numerical computations have not the logical trasparency of mathematical
derivations, but they can help understanding anyhow. Lattice can be used of
course to compute observable quantities (masses, matrix elements\ldots) from
first principles (Lattice for phenomenology),  
but  also as a tool to explore mechanisms
and structures of the theory by means of ``Gedanken experiments'' on the
configurations produced by numerical simulations (Lattice for theory).

We will be concerned with an investigation of the second type (Lattice for
theory), and test on the lattice the idea\cite{2,3,4} that confinement of colour is
produced by
{\em Dual superconductivity of type II of the QCD vacuum}.

The basic idea is that the chromoelectric field produced by a $Q \bar Q$ pair
is channeled by dual Meissner effect into Abrikosov flux tubes in the same way
as magnetic field is confined in usual superconductors of type II.

The energy is proportional to distance
\begin{equation}
E = \sigma R\label{eq:1.3}\end{equation}
and this means confinement. $\sigma$ is the string tension.

The world ``dual'' here means that the role of electric and magnetic quantities
is interchanged with respect to ordinary superconductors.

The idea is theoretically appealing in many respects.
\begin{itemize}
\item[1)] Superconductivity is a symmetry\cite{5}: deconfining transition is a change
of symmetry. An order (disorder) parameter can be defined and used to explore
superconductivity.
Dual superconductivity means a condensation
of monopoles in the ground state. Vacuum has no definite magnetic 
charge, but is a
superposition of states with different values of it\cite{6,6a}.

A disorder parameter for dual superconductivity will then be the 
vacuum expectation value (vev)
of any
operator carrying magnetic charge.
\item[2)] The very existence of monopoles implies that the theory is compact,
because of the Dirac quantization condition. Topology plays then an important
role, since Dirac strings make the connection of the gauge field non trivial.
A formulation in terms of parallel transport, like Wilson's lattice theory, is
then superior with respect to the perturbative formulation in terms of local
fields, which misses topology.
\item[3)] Monopole charges are always 
$U(1)$\cite{7,8}: in $SU(N)$ gauge theory there are
$N-1$ $U(1)$ magnetic charges. Monopole species are associated to any field
$\Phi$ in the adjoint representation 
\begin{equation}\Phi(x) = \sum_a\Phi^a\lambda^a
\label{eq:1a1}
\end{equation}
$\lambda^a$ are the generators of the gauge group in the fundamental
representation.

In field configurations monopoles are  located at the sites where two
eigenvalues of the $N\times N$ matrix $\Phi$, eq.(\ref{eq:1a1}),
 coincide. This is of course a
gauge invariant statement. The $N-1$ $U(1)$ fields associated with magnetic
charges are exposed by a gauge transformation $U(x)$ which diagonalizes
$\Phi(x)$. Such a transformation is called an abelian projection.

There are a functional infinity of choices for $\Phi$, and to each of them
monopole species are associated. What species do condense in the vacuum to
produce superconductivity is a dynamical question, and is actually what
has to be investigated to understand the mechanism of confinement. A
possibility is that all species are equally important for confinement and
condense in the vacuum\cite{9}.
\end{itemize}
One tool to investigate this issue is to define a disorder parameter
for different species and to measure it in connection with the deconfining
transition. As we shall see below this connection can be made unambigous by a
measurement of the critical indices or effective critical indices of the phase
transition, extracted from the behaviour of the disorder parameter.
A somewhat different attitude is to look at the abelian dominance\cite{10}. 
For a given abelian projection physical quantities as the string
tension are measured in the full theory and in the $U(1)$ theory resulting from
the 
abelian projection. If the latter determination is a good approximation to the
exact quantity, people say that there is abelian dominance. This happens to
80\% approximation in the so called maximal abelian projection. In addition one
can separate the abelian field into a part due to monopoles plus a residual
part with no topology\cite{11}. 
If the monopole part dominates people say that there is
monopole dominance. Again this happens in the maximal abelian projection.
Monopole dominance is then considered as a strong indication that confinement is
due to monopoles.

The two approaches are in our opinion both important to disantangle the
structure of the theory.

Our strategy is the following
\begin{itemize}
\item[a)] We define a disorder parameter for $U(1)$ dual superconductivity, and
test its construction in compact $U(1)$ gauge theory. The costruction is also
 tested with other well known systems, like the $XY$ 3d model, which
describes the transition to superfluid $He_4$.

In both cases we are able to detect the change of symmetry of the ground
state, and to determine the critical indices.
\item[b)] We  then define a similar disorder parameter for abelian
projected monopoles of non abelian gauge theories, and explore by it the
occurrence of dual superconductivity in connection with confinement.
\end{itemize}
A systematic analysis is in progress. Our preliminary results confirm that
monopoles defined by different abelian projection condense in the confined
phase, and the superconductivity disappears in the quark gluon phase,
supporting the view of ref.\cite{9}.

We will start these lectures by a brief introduction to basic superconductivity
in sect.2.

In sect.3 we shall recall the main properties of monopoles and the concept of
duality.

We will then describe the disorder parameter 
for condensation of monopoles
in $U(1)$ gauge theory and for
condensation of vortices in 3d $XY$ model (sect.'s 4,5).

The abelian projection and the physical meaning of monopoles will be 
discussed
in sect.6.
The results for $SU(2)$, $SU(3)$ and the other evidences from lattice for dual
superconductivity mechanism will be reviewed. 

Sect.7 will summarize
the state of the art
and present the open problems.

\section{Superconductivity as a symmetry.}
A relativistic version of the Ginzburg-Landau free energy, which is the
statistical analog of effective action, is
\begin{equation}
{\cal L} = -\frac{1}{4} F_{\mu\nu} F_{\mu\nu} + (D_\mu\Phi)^\dagger
(D_\mu\Phi) - V(\Phi)\label{eq:2.1}\end{equation}
where $\Phi$ is a complex (charged) scalar field describing Cooper 
pairs of charge $q= 2 e$.
\begin{equation}
D_\mu\Phi = (\partial_\mu - i q A_\mu)\Phi\label{eq:2.2}\end{equation}
is the covariant derivative,
\begin{equation}
V(\Phi) = \frac{\lambda}{4}(\Phi^\dagger\Phi -
\mu^2)\label{eq:2.3}\end{equation}
the potential, with $\mu$ and $\lambda$ functions of the temperature $T$, $\mu
= \mu(T)$, $\lambda = \lambda(T)$.

Minimizing ${\cal L}$ defines the ground state.
If $\mu^2>0$ the minimum corresponds to some $\langle\Phi\rangle\neq 0$, or to
the Higgs phase.
At $T$ where $\mu=0$ a transition to normal phase ($\mu^2< 0$, 
$\langle\Phi\rangle = 0$) takes place. $\langle\Phi\rangle$ is the order
parameter of superconductivity.

Putting $\Phi = \rho e^{i\theta}$, $\rho>0$,
under gauge transformations of angle $\alpha(x)$
\[\rho(x)\to \rho(x)\qquad \theta(x) \to \theta(x) + \alpha(x)
\qquad A_\mu(x)\to A_\mu(x) -
\partial_\mu \alpha\]
Therefore for the covariant derivative
\[ D_\mu\Phi =
e^{i\theta}\left[
\partial_\mu - i q(A_\mu - \partial_\mu\theta)\right]\rho\]
$\tilde A_\mu = (A_\mu - \partial_\mu\theta)$ is gauge invariant and 
$F_{\mu\nu} =
\partial_\mu \tilde A_\nu - \partial_\nu\tilde A_\mu$, since
$\partial_\mu\theta$ does not contribute.

${\cal L}$ can be rewritten as
\begin{equation}
{\cal L} = -\frac{1}{4} F_{\mu\nu} F_{\mu\nu} + \frac{m^2}{2}
\tilde A_\mu\tilde A_\mu + \tilde {\cal L}[\rho]
\label{eq:2.4}\end{equation}
and the equations of motion for the electromagnetic field read
\begin{equation}
\partial_\mu F_{\mu\nu} + m^2 \tilde A_\nu = 0
\label{eq:2.5}\end{equation}
with $m^2 = 2 q^2 \langle\Phi\rangle^2$.

In a stationary state with no charges $A_0=0$, $\partial_0\vec A = 0$ and 
equation (\ref{eq:2.5}) gives
$(\vec H = \vec\nabla\wedge\vec A)$
\begin{eqnarray}
\vec\nabla\wedge\vec H + m^2 \vec{\tilde A} &=& 0 \label{eq:2.6a}\\
\nabla^2 \vec H + m^2\vec H &=& 0\label{eq:2.6b}
\end{eqnarray}
Eq.(\ref{eq:2.6a}) means that a permanent current (London current) $\vec j = m^2
\vec{\tilde A}$ exists, and since $\vec E = 0$ and $\vec E = \rho \vec j$,
$\rho=0$. 
If $m^2$, or $\langle\Phi\rangle$ is
different from zero, there is superconductivity ($\rho=0$)

Eq.(\ref{eq:2.6b}) is nothing but Meissner effect: 
the penetration depth of the field $\vec H$ is $\lambda = 1/m$ 
and is again finite if
$\langle\Phi\rangle\neq 0$.

A side consequence of Meissner effect is flux quantization: outside a flux
tube, at distances larger than $\lambda$, $\vec{\tilde A}=0$. The integral
around a circle $C$ centered on the section of the tube of 
$\vec{\tilde A}$ is zero
\[0 = \oint\vec{\tilde A} \, d\vec x =
\oint(\vec A - \nabla \theta)  d\vec x\]
or, since the flux $F(H) = \oint \vec A  d\vec x$
\begin{equation}
F(H) = \frac{2\pi n}{q}\label{eq:2.7}\end{equation}
The key parameter is the order parameter $\langle\Phi\rangle$ which signals the
Higgs phenomenon.

$\langle\Phi\rangle\neq 0$ means condensation of charge. Indeed, if the ground
state has a definite charge, the expectation value on it of any charged
operator $C$ is zero: $\langle 0| C|0\rangle = 0$, since $C|0\rangle$ belongs to
a different eigenvalue of the charge than $|0\rangle$.

Superconducting vacuum is indeed known to be a coherent superposition of states
with different numbers of Cooper pairs\cite{6,6a}.

There are two characteristic lenghts in the system: 
the correlation length of the
$\Phi$ field, or the inverse Higgs mass $\Lambda = 1/M$, and the penetration
depth of the photon, $\lambda$. If $\lambda > \sqrt{2} \Lambda$ the
superconductor is called type II, and the formation of Abrikosov flux tubes is
favoured in the process of penetrating the material with a magnetic
field\cite{12}. If the opposite inequality holds, $\lambda < \sqrt{2}\Lambda$,
when the magnetic field is increased  there is an abrupt
penetration of it at some value and superconductivity is destroyed.

In principle many independent charged fields could condense in the vacuum. In
that case
\begin{equation}
\frac{m^2}{2} = \sum q_i^2 \langle \Phi_i\rangle^2
\label{eq:2.8}\end{equation}
\section{Monopoles and their topology.}
\subsection{$U(1)$ monopoles.}
Maxwell's equations in the presence of both electric and magnetic currents,
$j^\mu$, $j^\mu_M$, are
\begin{equation}
\partial_\mu F^{\mu\nu} = j^\nu\qquad \partial_\mu F^{*\mu\nu} = j^\nu_M
\label{eq:3.1}\end{equation}
$F_{\mu\nu}$ is the familiar field strength tensor, $F^*_{\mu\nu}$ its dual
\[ F^*_{\mu\nu} = \frac{1}{2}\varepsilon_{\mu\nu\rho\sigma}
F^{*\rho\sigma}\]
If both $j^\nu$ and $j^\nu_M$ are zero the transformation
\begin{eqnarray}
F_{\mu\nu} &\to& \cos\theta F_{\mu\nu} + \sin\theta F^*_{\mu\nu}
\label{eq:3.2a}\\
F^*_{\mu\nu} &\to& \cos\theta F^*_{\mu\nu} - \sin\theta F_{\mu\nu}
\label{eq:3.2b}\end{eqnarray}
is a symmetry of the system for any value of $\theta$. In particular for
$\theta=\pi$ the transformation becomes
\[ \vec E \to \vec H\qquad \vec H \to -\vec E\]
which is known as duality transformation.

In nature, within the present experimental limits, $j^\mu_M=0$, since no
isolated magnetic charge has been found. Therefore
\begin{equation}
\partial_\mu F^{*\mu\nu} = 0 \label{eq:3.3}\end{equation}
the general solution of this equation is
\begin{equation}
F_{\mu\nu} = \partial_\mu A_\nu - \partial_\nu A_\mu
\label{eq:3.4}\end{equation}
which makes eq.(\ref{eq:3.3}) identically satisfied (Bianchi identity).

Of course if $j^\mu_M\neq 0$ Bianchi identity is violated, and the vector
potential $A_\mu$ cannot be defined. The way out of this difficulty, was
proposed by Dirac\cite{13}. 
A monopole can be seen as the end point of a magnetic flux
tube, an infinitely thin solenoid (Dirac string), thus preserving
Bianchi identities. The solenoid will be 
 physically invisible if for any
particle with charge $e$ the parallel transport around it is trivial
\begin{equation}
\exp(i e \oint \vec A\,d\vec x) = 1\qquad
\hbox{or}\qquad \Phi\cdot e = 2\pi n
\label{eq:3.5}\end{equation}
The magnetic flux on the other hand is related to the visible magnetic charge
$M$ of the monopole as $\Phi = M$, and this requires
\begin{equation}
e M = 2\pi n\label{eq:3.6}\end{equation}
with $n$ integer, for any particle. Any charge is then multiple of the same
elementary charge
\begin{equation}
e = \frac{2\pi}{M}\label{eq:2.7a}\end{equation}
Theory is compact.
A formulation in terms of parallel transport is then superior in that it
naturally describes the non trivial connection of space time produced by the
presence of Dirac strings.

One could formulate the theory in a dual form, by introducing a dual vector
potential, and then $\partial_\mu F_{\mu\nu} = 0$ play the role of Bianchi
identities.
In this formulation charges would acquire a Dirac string of electric flux, and
monopoles would be pointlike.

Due to Dirac quantization condition, the weak coupling regime with respect to
charge corresponds to large values of $M$, or to strong coupling in the dual
language and viceversa. This is typical of systems with topological
excitations. 

The prototype model is the 2d Ising model. The field variable is
$\sigma(i)=\pm 1$ and the action
\[\beta J\sum_{i,\hat\mu}\sigma(i)\sigma(i+\hat \mu)\]
Looking at it as a $1+1$ dimensional field theory configurations like the one
in fig.1 are kinks and have a non trivial topology. One can define a dual
lattice, by associating a point $i'$ 
of it to each link of the original  lattice and
a field $\sigma^*(i') = \pm1$, assuming the value $+1$ if the sites at the ends
of the link have opposite sign, and  $-1$ if they have the same
sign. For a kink like the one in figure $\sigma^*$ is $+1$ at the position of
the kink, $-1$ everywhere else.
\vskip0.15in
\begin{minipage}{0.99\textwidth}
\epsfxsize = 0.99\textwidth
{\centerline{
{\epsfbox{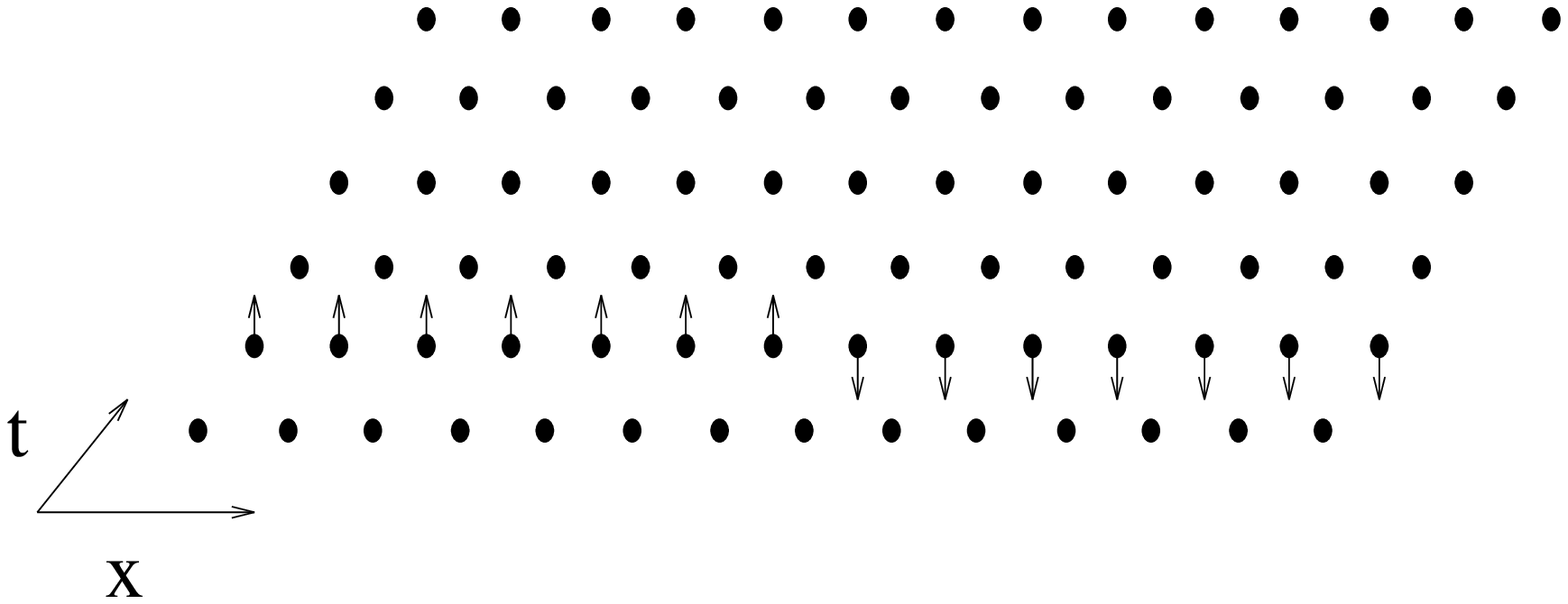}
}
}}
\par\noindent {\centerline{Fig.1\,
A kink in $2d$ Ising model.}}
\end{minipage}
\vskip0.15in\noindent
Duality in this case\cite{14} 
means that the partition function of the dual system has
the same form of the original one,
\begin{equation}
K[\sigma,\beta] = K[\sigma^*,\beta^*]\label{eq:3.10}\end{equation}
with the change $\beta\to\beta^*\sim1/\beta$.

The two systems are identical, but the low temperature (weak
coupling) regime of  one of  them corresponds to the high temperature
(strong coupling) of its dual. In the ordered phase $\langle\sigma\rangle \neq
0$,  $\langle\sigma^*\rangle=0$, and in the disordered phase
$\langle\sigma^*\rangle \neq
0$,  $\langle\sigma\rangle=0$. $\langle\sigma^*\rangle$ is called a disorder
parameter.
\subsection{Monopoles in non abelian gauge theories.}
Monopoles can exist as solitons, or static solutions with finite energy, in non
abelian gauge theories. They where discovered\cite{15,16} in the Georgi-Glashow
model, a gauge theory with gauge group $SO(3)$ coupled to a triplet of scalar
fields $\vec \phi$: the simplest generalization of the $U(1)$ Higgs model.
The lagrangean is
\begin{equation}
{\cal L} = -\frac{1}{4} \vec G_{\mu\nu} \vec G_{\mu\nu} +
(D_\mu\vec\phi)^\dagger (D_\mu\vec\phi) - V(\vec \phi)
\label{eq:3.11}\end{equation}
with $D_\mu$ the covariant derivative
\[ D_\mu\vec\phi = (\partial_\mu - g \vec A_\mu\wedge)\vec\phi\]
and $V(\vec\phi)$ the potential
\[V(\vec\phi) = \frac{\lambda}{4}\left(\vec\phi^2-\mu^2\right)^2\]
In the spontaneously broken phase, $\mu^2> 0$,
$\phi_0 = |\vec\phi_0| = \mu$. The ansatz
\begin{eqnarray}
\partial_0\vec\phi=0 &\quad& \vec\phi(\vec r) = f(r)\phi_0\,\hat r\qquad
\hat r=\frac{\vec r}{r}\label{eq:3.12a}\\
\vec A_0(\vec r) = 0&\quad& A_i^a(\vec r) =
h(r)\varepsilon_{iab}\frac{r_b}{g r^2}\label{eq:3.12b}
\end{eqnarray}
brings to a solution with finite energy, with
\[ h(r) \mathop\to_{r\to\infty} =1\qquad 
f(r) \mathop\to_{r\to\infty} =1
\]
The configuration is called an hedgehog, due to the form (\ref{eq:3.12a})
of the Higgs field.

That this configuration is  a monople can be seen by the following
arguments.
\begin{itemize}
\item[a)] At large distances, where $h(r) = f(r) = 1$ a gauge transformation
which brings $\hat \phi$ on some direction, say the 3 axis, 
(unitary gauge)
transforms the
gauge field to an abelian field, parallel to $\vec \phi$,  with a Dirac
string in space along the 3 axis.
\item[b)] A gauge invariant ``electromagnetic field'' can be defined\cite{15}
\begin{equation}
{\cal F}_{\mu\nu} = \hat\phi\cdot\vec G_{\mu\nu} -\frac{1}{g}
\hat\phi\left(D_\mu\hat\phi\wedge D_\nu\hat\phi\right)
\label{eq:3.13}\end{equation}
Here $\hat\phi \equiv \vec\phi/|\vec\phi|$.

${\cal F}_{\mu\nu}$, computed on the monopole configuration, is
\[ (\vec E)_i = {\cal F}_{0i} = 0\qquad
\vec H = \frac{1}{g}\frac{\hat r}{r^2}+\,\hbox{Dirac string}\]
$\vec H$ is a colour singlet, and such are the magnetic charges.
\end{itemize}

The gauge transformation bringing to the unitary gauge is called abelian
projection. 
For the monopole configuration 
it has a singularity at $\vec r=0$,
where $\hat\phi$ is not defined.

An alternative way to look at the problem is to use the Body Fixed Frame
(BFF)\cite{17}.
Usually the same reference frame for colour is used in all points of space
time, $\vec\xi_0^i$, with $\vec\xi_0^i\vec\xi_0^j = \delta^{ij}$,
$\vec\xi_0^i\wedge \vec\xi_0^j = \vec\xi_0^k$.
Instead three orthonormal unit vectors $\vec\xi_i(x)$ can be defined ( with
$\vec\xi^i\vec\xi^j = \delta^{ij}$, 
$\vec\xi^i\wedge \vec\xi^j = \vec\xi^k$) with $\vec\xi^3(x) = \hat\phi(x)$.
The choice of $\vec\xi_1(x)$, $\vec\xi_2(x)$ is arbitrary by an angle. The two
frames are related by a transformation of $SO(3)$, $R(x)$
\begin{equation}
\vec\xi_i(x) = R(x)\vec\xi^0_i\label{eq:3.14}\end{equation}
Since $(\vec\xi_i)^2 = 1$, $\partial_\mu\vec\xi_i$ is orthogonal to $\vec\xi_i$
and
\begin{equation}
\partial_\mu\vec\xi_i =
\vec\omega_\mu\wedge\vec\xi_i\label{eq:3.15}\end{equation}
or
\begin{equation}
(\partial_\mu - \vec\omega_\mu\wedge)\vec\xi_i \equiv D_\mu\vec\xi_i = 0
\label{eq:3.16}\end{equation}
Indeed the body fixed frame changes with $x$ by a parallel transport.

Eq.(\ref{eq:3.16}) also implies
\begin{equation}
\left[ D_\mu, D_\nu\right] \vec\xi_i = 0\label{eq:3.17}\end{equation}
From the completeness of $\vec\xi_i$, $\left[ D_\mu, D_\nu\right] = 0$.
This means
\begin{equation}
\vec G_{\mu\nu}(\omega)=
\partial_\mu\vec\omega_\nu - \partial_\nu\vec\omega_\mu +
\vec\omega_\mu\wedge\vec\omega_\nu = 0
\label{eq:3.18}\end{equation}
$\vec\omega_\mu$ is a pure gauge, at least in the regions of space where $R(x)$
is not singular.

The solution of Eq.(\ref{eq:3.16}) is
\begin{equation}
\hat\Phi(x) \equiv\vec\xi_3(x) =
P\,\exp\left[
i\int_{{\cal C}}^x \vec\omega_\mu\cdot\vec T\,d x^\mu\right] \vec\xi_0^3
\label{eq:3.19}\end{equation}
which is independent of the choice of the path ${\cal C}$ if $\vec
G_{\mu\nu}=0$. $P$ means path ordering and $\vec T$ are the generators of
$SO(3)$ group, $(T^j)_{ik} = - i\varepsilon_{ijk}$.

Expressing $\vec \xi_i(x)$ in terms of the polar angles $\theta$, $\psi$
with respect to $\vec\xi^0_i$, with polar axis along $\vec\xi^0_3$, one easily
finds
\begin{equation}
\vec\omega_\mu = \left(\matrix{
\sin\theta(x)\partial_\mu\psi(x)\cr
-\partial_\mu\theta(x)\cr
-\cos\theta(x)\partial_\mu\psi(x)\cr}\right)
\label{eq:3.20}\end{equation}
At $\theta=0,\pi$, $\psi$ is not defined and a singular part of
$\vec\omega_\mu$ develops
\begin{equation}
\vec\omega_\mu^{sing} = \left(\matrix{
0\cr
0\cr
\pm\partial_\mu\psi^{sing}(x)\cr}\right)
\label{eq:3.21}\end{equation}
and with it a singular field strength tensor
\begin{equation}
\vec F_{\mu\nu}(\omega) = \pm(\partial_\mu\partial_\nu-
\partial_\nu\partial_\mu)\psi^{sing} \vec\xi_3(x)
\label{eq:3.22}\end{equation}
$\vec F_{\mu\nu}$ is abelian and parallel to $\hat\phi = \vec\xi_3$.

For the static monopoles $\vec F_{\mu\nu}(\omega)$ is a Dirac string along the
3 axis at all times: the end point of the string is the location of the
monopole, i.e. the zero of $\vec\phi(x)$.

The monopole is not an artefact of the abelian projection, but a topological
feature of the $\vec\phi$ field configuration.

The monopole configuration described above
is also a soliton, and behaves like a particle.

However  singularities in $\vec \omega_\mu$ can exist also in the unbroken
phase of the theory, 
at the zeros of $\vec \phi$,
and they are monopoles. 
Again they are topological properties of the $\vec\phi$ field configurations.

Under infinitesimal gauge transformations $\exp(i\vec\lambda(x)\vec T)$
\begin{eqnarray*}
\vec\omega_\mu &\to& \vec\omega_\mu + \vec\lambda\wedge\vec\omega_\mu +
\partial_\mu\vec \lambda\\
\vec A_\mu &\to& \vec A_\mu + \vec\lambda\wedge\vec A_\mu -\frac{1}{g}
\partial_\mu\vec \lambda
\end{eqnarray*}
so that
\begin{equation}
\vec\omega_\mu + g\vec A_\mu = g \vec Z_\mu \label{eq:3.23}\end{equation}
is covariant.

In the abelian projected gauge $\vec\omega_\mu=0$ and $\vec A_\mu = \vec
Z_\mu$. The field strength tensor can be computed, obtaining
\begin{equation}
\vec G_{\mu\nu}(Z) = \vec G_{\mu\nu}(A) + \frac{1}{g} \vec F_{\mu\nu}(\omega)
\label{eq:3.25}\end{equation}
The gauge transformation which operates the abelian projection is singular, and
that produces the additional term in $\vec G_{\mu\nu}$.
Also
\begin{equation}
D_\mu(A)\hat \phi = (\vec\omega_\mu + g \vec A_\mu)\wedge\vec\varphi =
g \vec Z_\mu \wedge\hat\varphi\label{eq:3.26}\end{equation}
and as a consequence
\begin{equation}
\frac{1}{g}\hat\varphi(D_\mu\hat\varphi\wedge D_\nu\hat\varphi) =
g(\vec Z_\mu\wedge \vec Z_\nu)\cdot\hat\varphi
\label{eq:3.27}\end{equation}
Thus the t'Hooft e.m. gauge invariant tensor
\[
{\cal F}_{\mu\nu} = \vec\phi\cdot\vec G_{\mu\nu} -\frac{1}{g}
\hat\phi\left(D_\mu\hat\phi\wedge D_\nu\hat\phi\right)
\]
reads in the abelian projected gauge
\begin{equation}
{\cal F}_{\mu\nu} = \partial_\mu Z^3_\nu - \partial_\nu Z^3_\mu
\label{eq:3.28}\end{equation}
since the non abelian term in $\vec\phi\cdot\vec G_{\mu\nu}$ is canceled by the
additional term by virtue of eq.(\ref{eq:3.27}).

The gauge invariant tensor which describes  the $U(1)$ field coupled to monopole
charge coincides with the abelian field of the residual $U(1)$ after abelian
projection.  Because of that sometimes instead of associating monopoles to the
field $\vec \phi$, people associate them to the gauge which puts $\hat \phi$
along the 3 axis.

Since in QCD the possible fields $\hat \phi$ which could define monopoles
are many, and to each of them a different gauge transformation is associated as
abelian projection, the statement could be made  that monopoles are gauge
dependent objects or gauge artefacts.
This is a misuse of language. 

The physical problem is to identify what
$\vec\phi(x)$ fields 
are relevant to confinement, in that  monopoles associated to
them 
condense and produce dual superconductivity.
Monopole species 
depend on the choice of $\vec\phi$. However the monopole charges and
electromagnetic
field associated to each $\vec \phi$ are gauge invariant concepts.
\section{A disorder parameter for dual superconductivity: compact $U(1)$ gauge
theory.}
In any field theory in which non trivial topological objects $T$ exist,
like monopoles or vortices, with a conserved topological charge, a creation
operator can be defined for them.

The original construction goes back to ref.\cite{14}, and has been developed in
different forms by many authors\cite{18,18a,18b,20,b17}. 
The basic idea is translation, in
the sense of the elementary formula
\begin{equation}
e^{i p a}|x\rangle = |x + a\rangle\label{eq:4.1}\end{equation}
If $\Phi(x)$ is the field describing the theory and $\Phi_T(x,y)$ is the field
configuration  produced by,say,  a monopole sitting at $y$, then the operator
\begin{equation}
\mu(\vec y,t) = \exp\left(
i \int d^3x\,\Pi_\Phi(\vec x,t)\Phi_T(x,y)\right) \label{eq:4.2}\end{equation}
with $\Pi_\Phi(\vec x,t)$ the conjugate momentum to $\Phi(\vec x,t)$, creates a
monopole at the site $\vec y$ and time $t$.

Indeed in the Schr\"odinger representation $|\Phi(\vec x)\rangle$
\[\mu(\vec y,t) |\Phi(\vec x)\rangle =
|\Phi(\vec x) + \Phi_T(\vec x,\vec y)\rangle\]
We will focus on compact $U(1)$ gauge theory in 4d on lattice\cite{22,23}, 
where monopoles
exist and magnetic charge is conserved.

The building block of the theory is the parallel transport along links of the 
lattice, exiting from site $n$ in direction $\mu$
\begin{equation}
U_\mu(n) = \exp(i e a A_\mu(n))\equiv
\exp(i \theta_\mu(n))\label{eq:4.3}\end{equation}
The parallel transport along the plaquette, the elementary square in the plane
$\mu\nu$, is then
\[ \Pi_{\mu\nu} = \exp(i \theta_{\mu\nu}(n))\]
\begin{equation}
\theta_{\mu\nu}(n) =
\Delta_\mu \theta_\nu - \Delta_\nu\theta_\mu\simeq a^2 e f_{\mu\nu}
\label{eq:4.4}\end{equation}
The generating functional of the theory or partition function is
\begin{equation}
Z(\beta) = \int\prod_{n,\mu}\left(\frac{d\theta_\mu(n)}{2\pi}\right)
\exp(-S) \label{eq:4.5}\end{equation}
We will choose for $S$ the Wilson action
\begin{equation}
S = \beta\sum_{n,\mu<nu}\left(1 - \cos\theta_{\mu\nu}(n)\right)
\label{eq:4.6}\end{equation}
As $\beta\to\infty$ small values of $\theta_{\mu\nu}$ are important and
\begin{equation}
S\mathop\simeq_{\beta\to\infty}\frac{1}{4}\beta\sum \theta_{\mu\nu}^2 =
\beta a^4\sum e^2 f^2_{\mu\nu}\label{eq:4.7}\end{equation}
which is the action for free photons if the identification is made
$\beta=1/e^2$.
Compactness of the theory, i.e. the fact that angles $\theta_\mu(n)$ only
appear as arguments of periodic functions in the action, makes $Z(\beta)$
and correlations functions of compact field variables
invariant under the change
\[\theta_\mu(n) \to \theta_\mu(n) + f_\mu(n)\]
with arbitrary $f_\mu(n)$. A special case are gauge transformations $f_\mu(n) =
\Delta_\mu\Phi$.

A critical $\beta_c$ exists in the model, 
$\beta_c\simeq 1.0116$\cite{24,24a}. 

For $\beta >
\beta_c$ the theory describes free photons. For $\beta < \beta_c$ electric
charge is confined.
Wilson loops obey the area law, and dual Meissner effect is observed.

Monopoles exist in this theory. Indeed, since
\[-\pi \leq \theta_\mu(n) \leq \pi\]
from eq.(\ref{eq:4.4})
\begin{equation} -4\pi \leq \theta_{\mu\nu}(n)\leq
4\pi\label{eq:4.6a}\end{equation}
Since in the  plaquette integer multiples of $2\pi$ are not visible,
because of compactness, one can redifine $\theta_{\mu\nu}$ as
\begin{equation}
\theta_{\mu\nu} = 2\pi n_{\mu\nu} + \bar\theta_{\mu\nu}
\quad -\pi \leq \bar\theta_{\mu\nu}\leq \pi
\label{eq:4.7a}\end{equation}
$\bar\theta_{\mu\nu}$ is the visible flux.

Now 
\[\theta^*_{\mu\nu} =
\frac{1}{2}\varepsilon_{\mu\nu\rho\sigma}\theta_{\mu\nu}\]
obeys the Bianchi identities
$\Delta_\mu\theta^*_{\mu\nu} = 0$ as can be trivially checked. The visible
field, however, is $\bar\theta_{\mu\nu}$, because of compactness and can
violate Bianchi identities because units of $2\pi$ can be formed which becoomes
invisible. A monopole current can be defined as
\begin{equation}
\rho^M_\mu = -\frac{1}{6}\varepsilon_{\mu\nu\rho\sigma}\Delta_\nu n_{\rho\sigma}
\label{eq:4.8}\end{equation}
and
\begin{equation}
\Delta_\mu\bar\theta_{\mu\nu} = \rho^M_\mu\label{eq:4.9ma}\end{equation}
Monopoles are identified as Dirac strings. Since $\rho^M_\mu$ is 
identically
conserved,
(see eq.(\ref{eq:4.8})) strings are closed.

In the pioneering work of ref.(\cite{22}) such monopoles where 
numerically detected.
Their density being higher in the confined phase, $\beta < \beta_c$ and
dropping to zero above $\beta_c$, the density of monopoles was called an order
parameter for the transition. Although the observed correlation between density
and phase is phenomenological significant, a genuine disorder parameter for the
system should be related to the symmetry of the ground state
and should then be the vev of a magnetically charged operator.
A candidate
disorder parameter is the vacuum expectation value of the creation operator of
a monopole.

In the continuum the general rule eq.(\ref{eq:4.2}) gives for that operator
\begin{equation}
\mu(\vec y,t) =
\exp\left[i \int d^3x\,\vec E(\vec x,t)\frac{1}{e}\vec b(\vec x - \vec y)\right]
\label{eq:4.9}\end{equation}
$\vec E(\vec x,t)$ being the conjugate momentum to the field $\vec A(\vec x,t)$,
and $1/e\,\vec b$ the 
classical configuration corresponding to a monopole sitting at $\vec y$
\begin{equation}
\vec A_{mon}(\vec x-\vec y) = \frac{1}{e}\vec b(\vec x -\vec y)\label{eq:4.10}
\end{equation}
the factor $1/e$ coming from Dirac quantization condition is explicitely
exposed. With some choice of the classical gauge, putting the string along the
unit vector $\vec n$
\begin{equation}
\frac{1}{e}\vec b(\vec r) = \frac{2\pi m}{e}\frac{\vec r\wedge\vec n}{r(r-\vec
r\cdot \vec n)}
\label{eq:4.11}\end{equation}
A change of the classical gauge is reabsorbed in the definition (\ref{eq:4.9}) 
if $\vec E$
obeys Gauss's law.

On the lattice 
\[\Pi_i = \frac{1}{e}{\rm Im}\Pi^{0i} = \frac{1}{e}\sin\theta^{0i}\]
so that the naive transcription of eq.(\ref{eq:4.9}) would be\cite{18a}
\begin{equation}
\mu(\vec y,n_0) =\exp\left[\beta\sum_{\vec n}
b^i(\vec n-\vec y)\sin\theta^{i0}(\vec n,n_0)\right]
\label{eq:4.12}\end{equation}
The factor $\beta$ comes from the $1/e$ in the magnetic charge times $1/e$ in
the normalization of the plaquette to the electric field.

The above definition can be adapted to compactness, i.e. to give a shift of the
angle instead of its sinus, as follows\cite{a17}
\begin{equation}
\mu(\vec y,n_0) =
\exp\left[\tilde S(\theta^{i0}(n_0) + b^i(\vec n-\vec y)) -
\tilde S(\theta^{i0}(n_0))\right]\label{eq:4.13}\end{equation}
$\tilde S$ is the sum of the density of action on the time slice $n_0$
\begin{equation}
\tilde S = \sum_{\vec n,\mu,\nu} {\cal L}[\theta_{\mu\nu}(\vec n,n_0)]
\label{eq:4.14}\end{equation}
Since in the limit $a\to 0$, ${\cal L}\to
\frac{\beta}{2}\sum_{\mu<\nu}\theta_{\mu\nu}^2$
\begin{equation}
\mu\mathop\simeq_{a\to 0}
\sum_{\vec n} \theta^{i0}(\vec n,n_0) b_i(\vec n-\vec y) \beta +
\sum_{\vec n} b_i^2(\vec n-\vec y)\end{equation}
and $\mu$ coincides with the naive definition modulo a constant coming from the
last term.

If more monopoles or antimonopoles are created at time $n_0$, $b^i$ should be
replaced by the classical field describing their configuration.

Correlation functions of monopoles and or monopoles antimonopoles can then be
constructed.

We will focus on the correlation\cite{a17}
\begin{equation}
{\cal D}(x_0) =
\langle \bar \mu(\vec 0,x_0)\,\mu(\vec 0,0)\rangle 
\label{eq:4.14a}\end{equation}
between a monopole sitting at $\vec 0$ in space at time $0$, and an
antimonopole at $\vec 0$ and time $x_0$ (propagator of monopole field).

At large $x_0$ we expect, by cluster property
\begin{equation}
{\cal D}(x_0)\simeq A\exp(- M |x_0|) + \langle \mu\rangle^2 
\label{eq:4.15}\end{equation}
Translation and $C$ invariance make
\begin{equation}
\langle\bar \mu(\vec 0,x_0)\rangle =
\langle\mu(\vec0,0)\rangle\equiv\langle\mu\rangle
\label{eq:4.15a}\end{equation}
$\langle\mu\rangle$ is our disorder parameter: $\langle\mu\rangle\neq 0$
signals spontaneous breaking of magnetic $U(1)$ and hence dual
superconductivity, as discussed in sect.2.

The other important quantity in eq.(\ref{eq:4.15}) is $M$, which is the lowest
mass of excitations carrying monopole charge. The effective scalar field
producing dual superconductivity has  a mass larger or equal to $M$, and
hence knowledge of $M$ is an important information to determine the type of dual
superconductivity.

Before going to numerical results we will clarify our definition of $\mu$, by
analyzing in detail ${\cal D}(x_0)$. According to that definition 
eq.(\ref{eq:4.15})
\begin{equation}
{\cal D}(x_0) =
\frac{1}{Z[S]} Z[S+\Delta S]\label{eq:4.16}\end{equation}
The factor $1/Z[S]$ comes from the averaging procedure, $Z[S+\Delta S]$ is
nothing but the partition function with a modified action: using the definition
eq.(\ref{eq:4.13}) the modification consists in replacing
\begin{eqnarray}
\tilde S(\theta^{i0}(0)) &\to&
\tilde S[\theta^{i0}(0) + b^i(\vec y)] \label{eq:4.17a}\\
\tilde S(\theta^{i0}(n_0)) &\to&
\tilde S[\theta^{i0}(n_0) - b^i(\vec y)]\label{eq:4.17b}\end{eqnarray}

Since 
\[ \theta^{i0}(\vec n,0) = -\theta^i(\vec n,1) + \theta^i(\vec n,0) +
\theta^0(\vec n+\hat i,0) - \theta^0(\vec n,0)\]
the change implied by eq.(\ref{eq:4.17a}) can be reabsorbed by a change of
integration
variables 
\begin{equation}
\theta^i(\vec n,1)\to \theta^i(\vec n,1) + b^i(\vec n)\label{eqn1}
\end{equation}
which leaves $Z$ unchanged, because of compactness.

The result of this change is that $\theta^{i0}(0) $ is restored to the form it
has in $Z[S]$, but at $n_0=1$ 
\[\theta^{ij}(\vec n,1)\to \theta^{ij}(\vec n,1) + \Delta_i b_j - \Delta_j b_i
\]
A monopole field is added at $n_0=1$, in a form which is independent of the
gauge choice for $b_i$.  $\theta^i(\vec n,1)$ also appears in
$\theta^{i0}(\vec n,1)$ and, the change of variables gives
\begin{equation}
\theta^{i0}(\vec n,1)\to \theta^{i0}(\vec n,1) + b_i(\vec n)
\label{eq:4.20}\end{equation}
which is the same as (\ref{eq:4.17a}), at $n_0=1$.
We can now repeat the procedure, and the result will be a monopole at time
$n_0=2$ and again a change  of the form
(\ref{eq:4.20})
 at time 2. The procedure
ends at $n_0-1$ where the change (\ref{eqn1}) is reabsorbed by the
antimonopole, eq.(\ref{eq:4.17b}).
Our construction really produces a monopole at site $\vec 0$, propagating from
0 to $x^0$.

A direct determination of $\mu$ from eq.(\ref{eq:4.15}) 
at large value of $x_0$ is
shown in fig.2.
\vskip0.1in
\begin{minipage}{0.99\textwidth}
\epsfxsize = 0.99\textwidth
{\centerline{
{\epsfbox{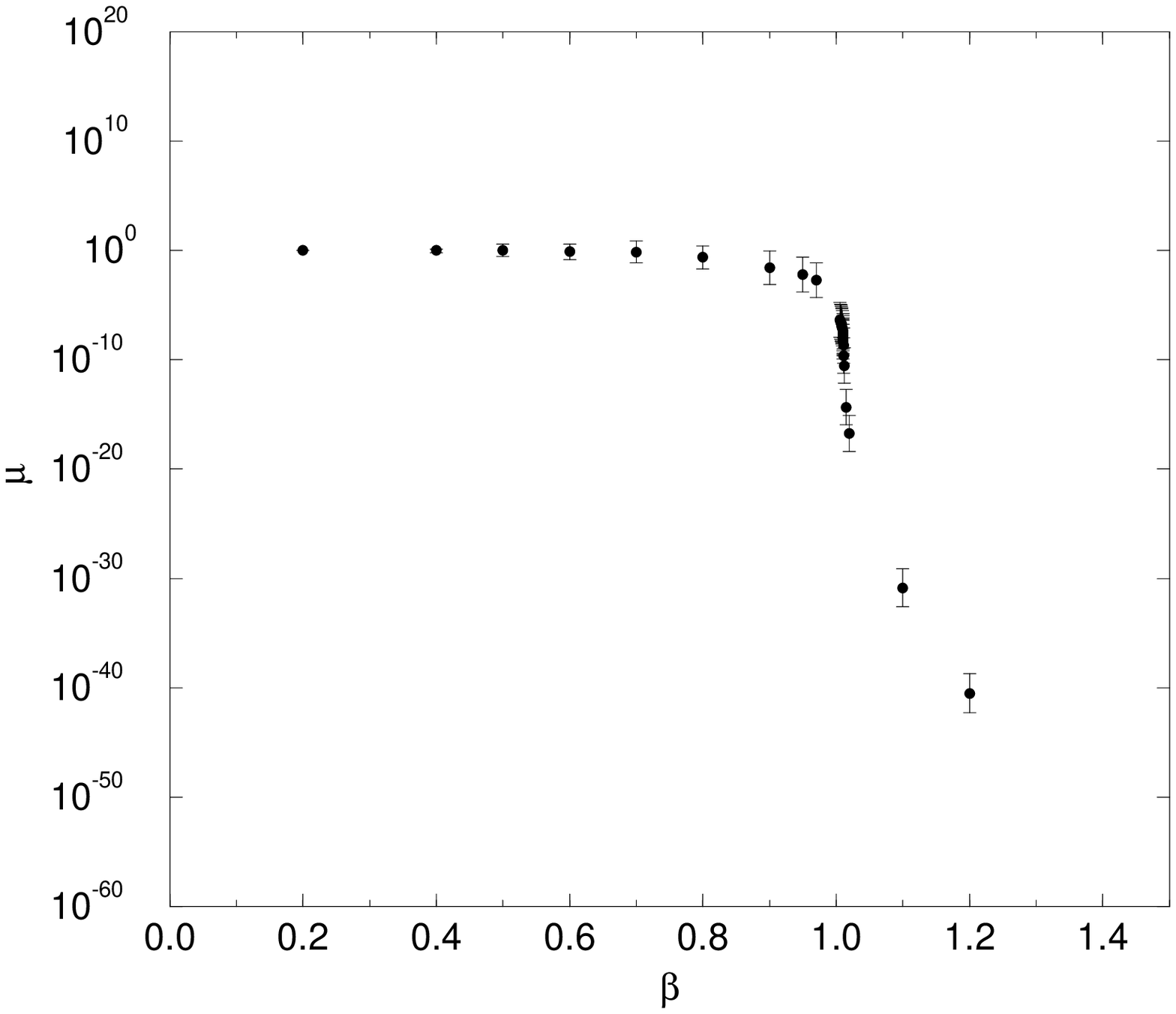}
}
}}
\vskip-0.05in
\par\noindent {\centerline{ Fig.2\,
The disorder parameter vs $\beta$ on a $10^4$ lattice.}}
\end{minipage}
\vskip0.15in\noindent
Instead of ${\cal D}(x_0)$ itself it provides numerically convenient to compute
\begin{equation}
\rho(x_0) = \frac{d}{d\beta}\ln {\cal D}(x_0)
\label{eq:4.22}\end{equation}
or, by eq.(\ref{eq:4.16})
\[ \rho = \langle S\rangle_S - \langle S+\Delta S\rangle_{ S+\Delta S}\]
The subscript of the brakets denotes the action used in performing average.
Again as $|x_0|\to \infty$
\begin{equation}
\rho(x_0) \mathop\simeq_{|x_0|\to \infty} 2 \frac{d}{d\beta}
\ln\langle\mu\rangle + C \exp(-M |x_0|)
\label{eq:4.23}\end{equation}
The typical behaviour of $\rho(x_0)$ near the transition is shown in fig.3.
The typical correlation length is of the order of the lattice spacing and  
$M$ can be determined.

The behaviour of $\rho_\infty$ is shown in fig.4. 
The sharp drop of $\langle\mu\rangle$
around $\beta_c$ is reflected in a narrow negative peak in $\rho$.

\vskip0.1in\noindent
\begin{minipage}{0.99\textwidth}
\epsfxsize = 0.99\textwidth
{\centerline{
{\epsfbox{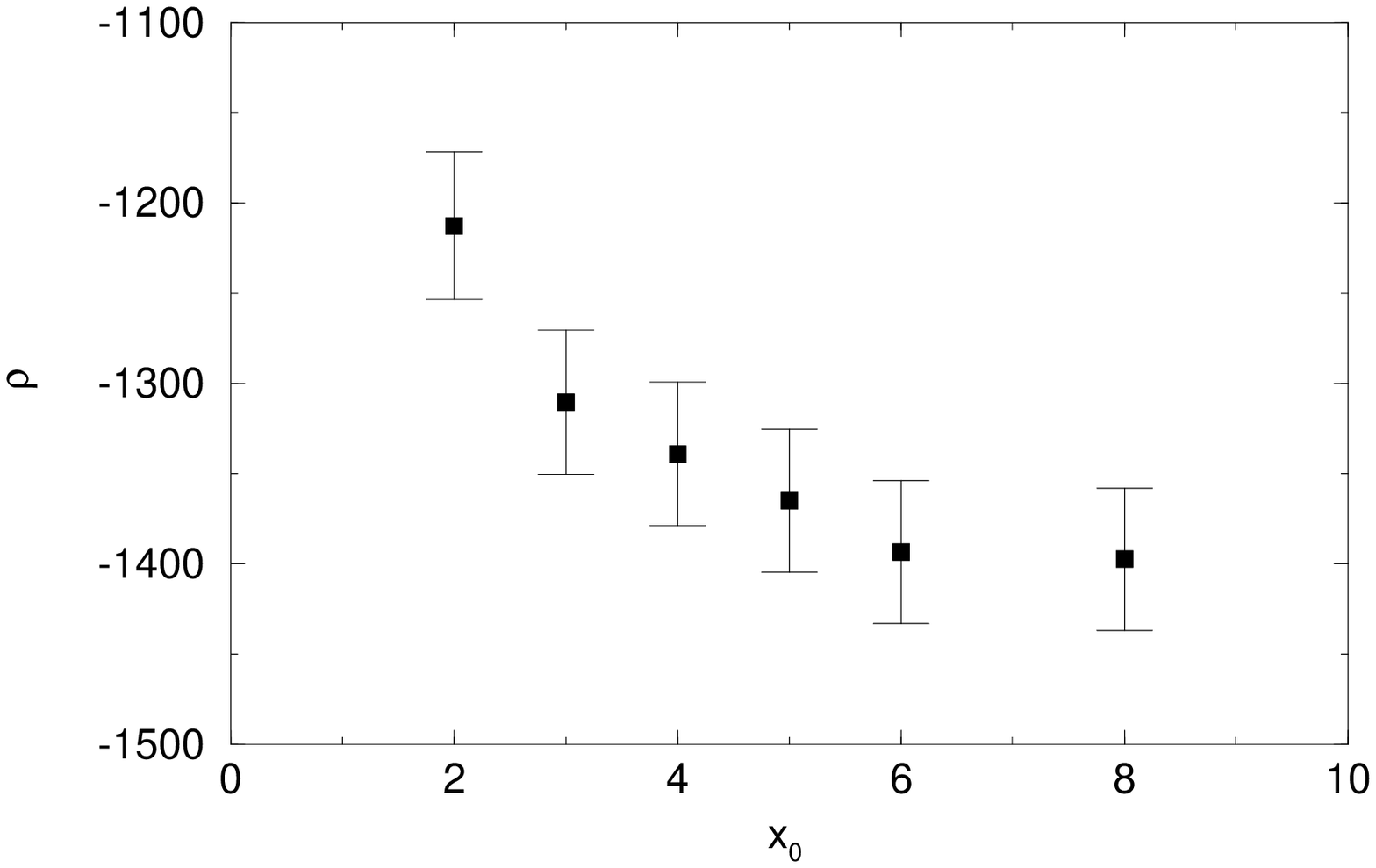}
}
}}
\par\noindent {\centerline{ Fig.3\,
$\rho(x_0)$ vs $x_0$ at $\beta = 1.099$. The correlation length is of the order
of lattice spacing.}}
\end{minipage}
\vskip0.1in\noindent
\begin{minipage}{0.99\textwidth}
\epsfxsize = 0.99\textwidth
{\centerline{
{\epsfbox{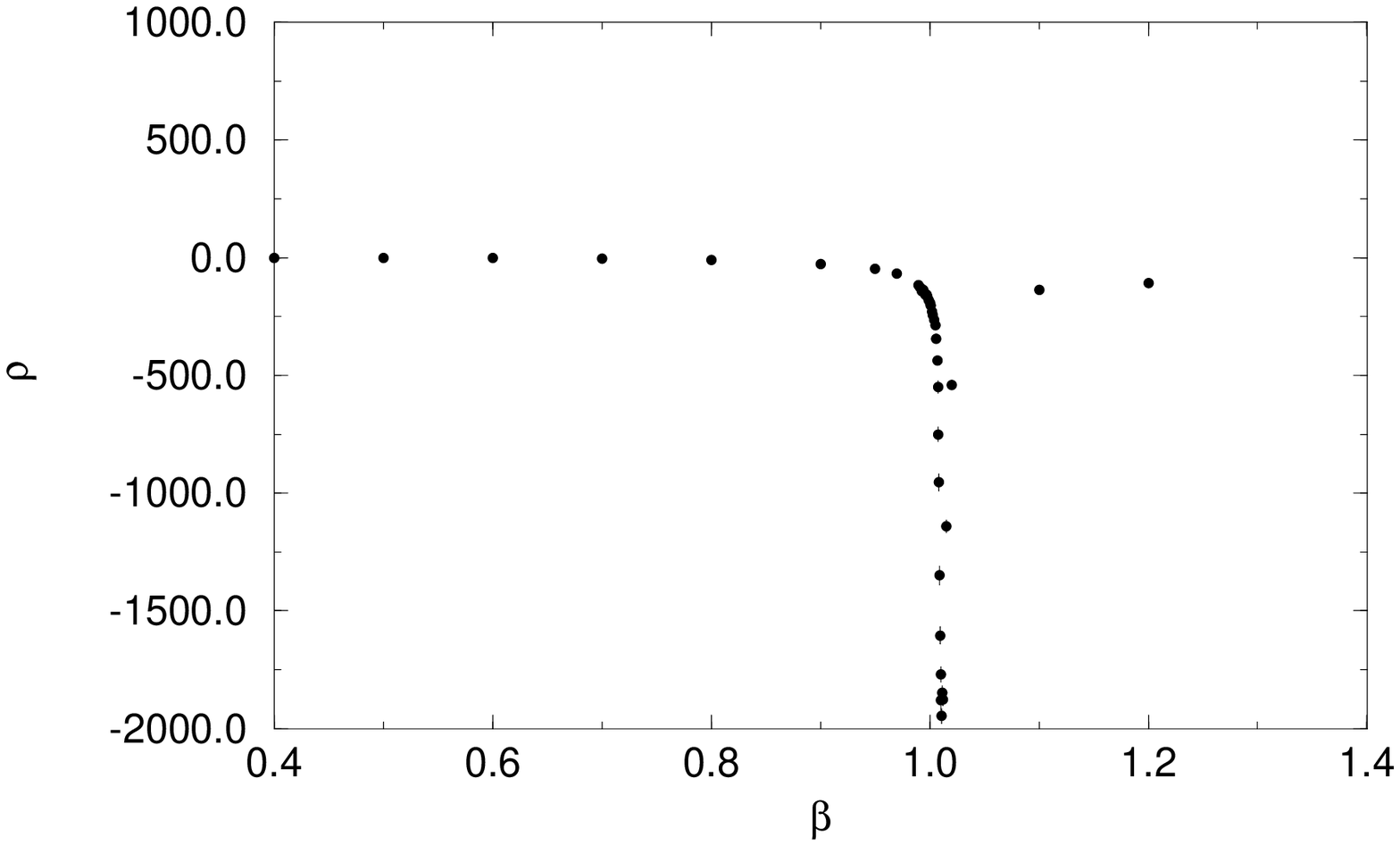}
}
}}
\par\noindent {\centerline{ Fig.4\,
$\rho$ as a function of $\beta$. The sharp negative peak indicates the phase 
transition. }}
\end{minipage}
\vskip0.15in\noindent

For $\beta > \beta_c$ the system describes free photons, the Feynman integral
is gaussian and $\rho$ can be explicitely computed. 
Numerically one finds for a lattice
$L^3\times 2 L$
\begin{equation}
\rho_\infty = -10.1\,L + 9.542 \label{eq:4.24}\end{equation}
$\rho_\infty$ tends to $-\infty$ as the volume goes large (thermodynamical
limit). and
\[ \mu = \exp\left(\int_0^\beta \rho(x)\,d x\right)\]
tends to zero. Above $\beta_c$ magnetic $U(1)$ is not broken and $\langle
\mu\rangle= 0$. Notice that this can only be true in the thermodynamical 
limit\cite{14}. In
a finite volume $\langle\mu\rangle$ is an analitic function of $\beta$ and
cannot be identically zero on a line of the complex plane without being zero
everywhere. Only as $L\to\infty$ Lee-Yang singularities develop and
$\langle\mu\rangle$ can be zero.

For $\beta < \beta_c$ $\rho_\infty$ tends to a finite value, compatible with
zero as $L\to \infty$, and hence $\langle\mu\rangle\neq 0$ and the system is a
dual superconductor.

The behaviour of $\langle\mu\rangle$ around $\beta_c$ can be explored by a
finite size scaling analysis. At $\beta\simeq \beta_c$ a weak first order or
second order phase transition takes place. The order is 
controversial\cite{24,24a}.
In any case
the correlation length $\xi$ goes large in a range of values
around $\beta_c$, and an effective critical index $\nu$ can be  defined
\begin{equation}
\xi \mathop\simeq_{\beta\to\beta_c} (\beta_c - \beta)^{-\nu}
\label{eq:4.25}\end{equation}
By dimensional analysis
\[\langle\mu\rangle = \mu(\frac{L}{\xi},\frac{a}{\xi})\]
as $\xi\to large$ the dependence on $a/\xi$ can be neglected and
\begin{equation}
\langle\mu\rangle = \mu(\frac{L}{\xi},0) =
f(L^{1/\nu}(\beta_c-\beta))\label{eq:4.26}\end{equation}
By use of eq.(\ref{eq:4.25}) the variable $L/\xi$ has been traded with
$L^{1/\nu}(\beta_c-\beta)$. 

The following 
scaling law follows for $\rho = \frac{d}{d\beta}\ln\langle\mu\rangle$
\begin{equation}
\frac{\rho}{L^{1/\nu}} = - \frac{f'}{f} = \Phi(L^{1/\nu}(\beta_c-\beta))
\label{eq:4.27}\end{equation}
This scaling can be matched by appropriate values of $\nu$ and $\beta_c$. We
find by best fit
\begin{equation}
\beta_c = 1.01160(5)\qquad \nu = 0.29(2)\label{eq:4.28}\end{equation}
$\beta_c$ agrees with determinations based on completely different 
methods\cite{24}.
The
quality of scaling is shown in fig.5. 
\vskip0.1in\noindent
\begin{minipage}{0.99\textwidth}
\epsfxsize = 0.99\textwidth
{\centerline{
{\epsfbox{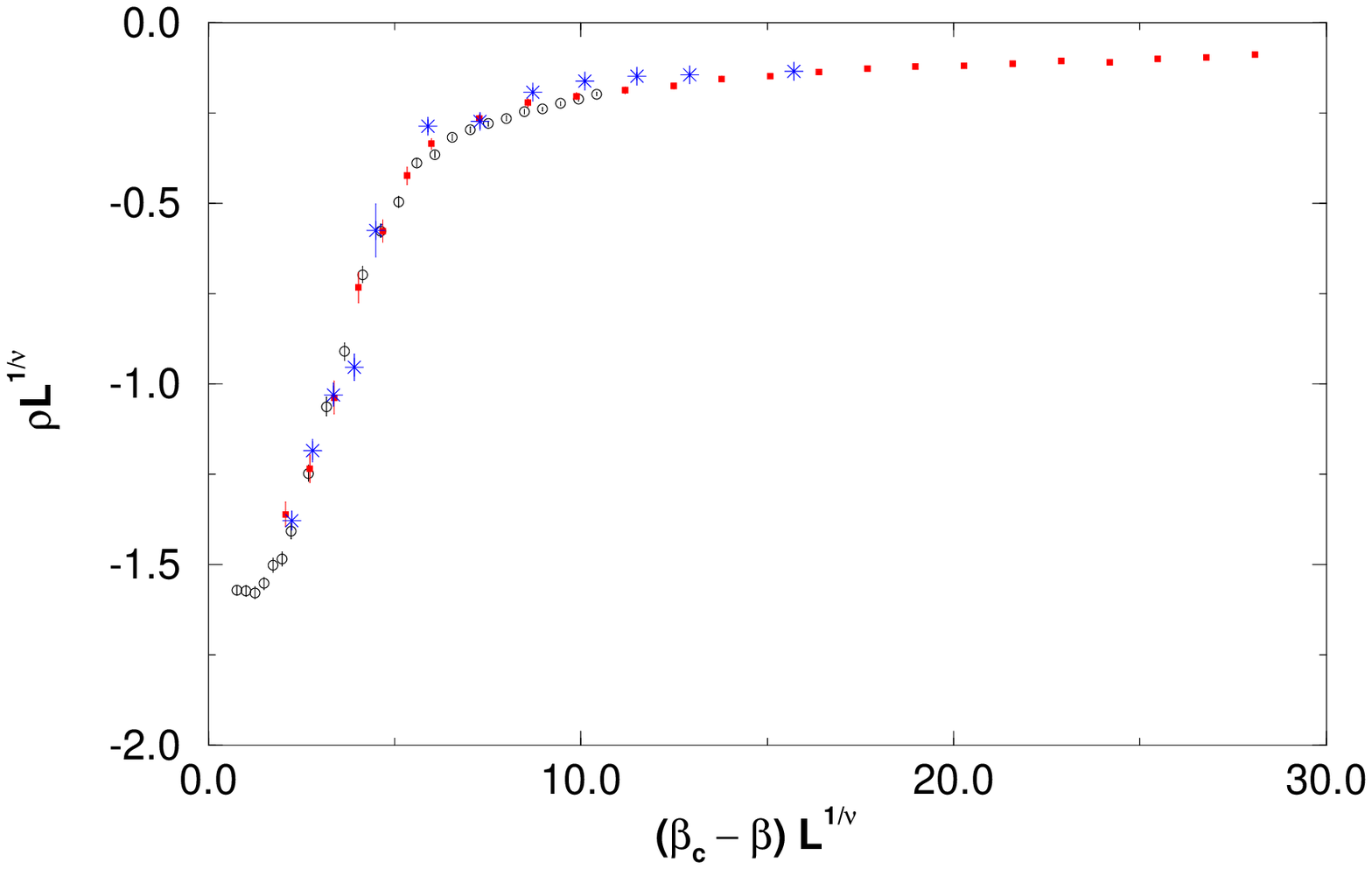}
}
}}
\vskip0.05in\noindent {\centerline{ Fig.5\,
Finite size scaling of $\rho$, at the optimal values of $\nu$, $\beta_c$.
}}
\end{minipage}
\vskip0.15in\noindent
We can also compute the critical
index by which $\mu\to 0$ at $\beta_c$
$\langle\mu\rangle\sim (\beta_c-\beta)^\delta$ finding
\begin{equation}
\delta = 1.1\pm0.2\label{eq:4.29}\end{equation}
A first order transition would require $\nu = 1/d = 0.25$. Further
investigation is on the way to test if the observed value of $\nu$ is a finite
size effect or it is a true determination and the transition is 2nd order.

We have also measured the penetration depth of the field $E$, or the  mass $m$
of the photon. The result is shown in fig.6, where $m$ is compared to $M$.

Further work is necessary to reduce the errors in $M$. There seems however to
be sufficient information, at least in the region near $\beta_c$, to
conclude that the ratio $M/m > \sqrt{2}$, i.e. that the superconductor is
type II.

Similar results was found by direct observation of the London current in a 
flux
tube, in ref.(\cite{26}).

If the transition is first order 
the lattice model does not define a field theory as
$\beta\to \beta_c$ and the only fixed point is the trivial point
$\beta=\infty$, which describes free photons. If this is the case changing the
lagrangian from Wilson to an alternative form, say Villain, also changes the
physics of the system, since there is no universality class.

For Villain action the duality transformation can be performed, and 
condensation
of monopoles below the critical value has been proved\cite{20}. 
In that case it can be
shown as a theorem that our disorder parameter is equal to that of 
ref.(\cite{20}),
although the construction is completely different.
Both constructions are based on eq.(\ref{eq:4.2}).
The proof of ref.(\cite{20}) 
has been extended to Wilson action in ref.(\cite{25}).

In conclusion we have a reliable tool to detect dual superconductivity.
\vskip0.1in\noindent
\begin{minipage}{0.99\textwidth}
\epsfxsize = 0.99\textwidth
{\centerline{
{\epsfbox{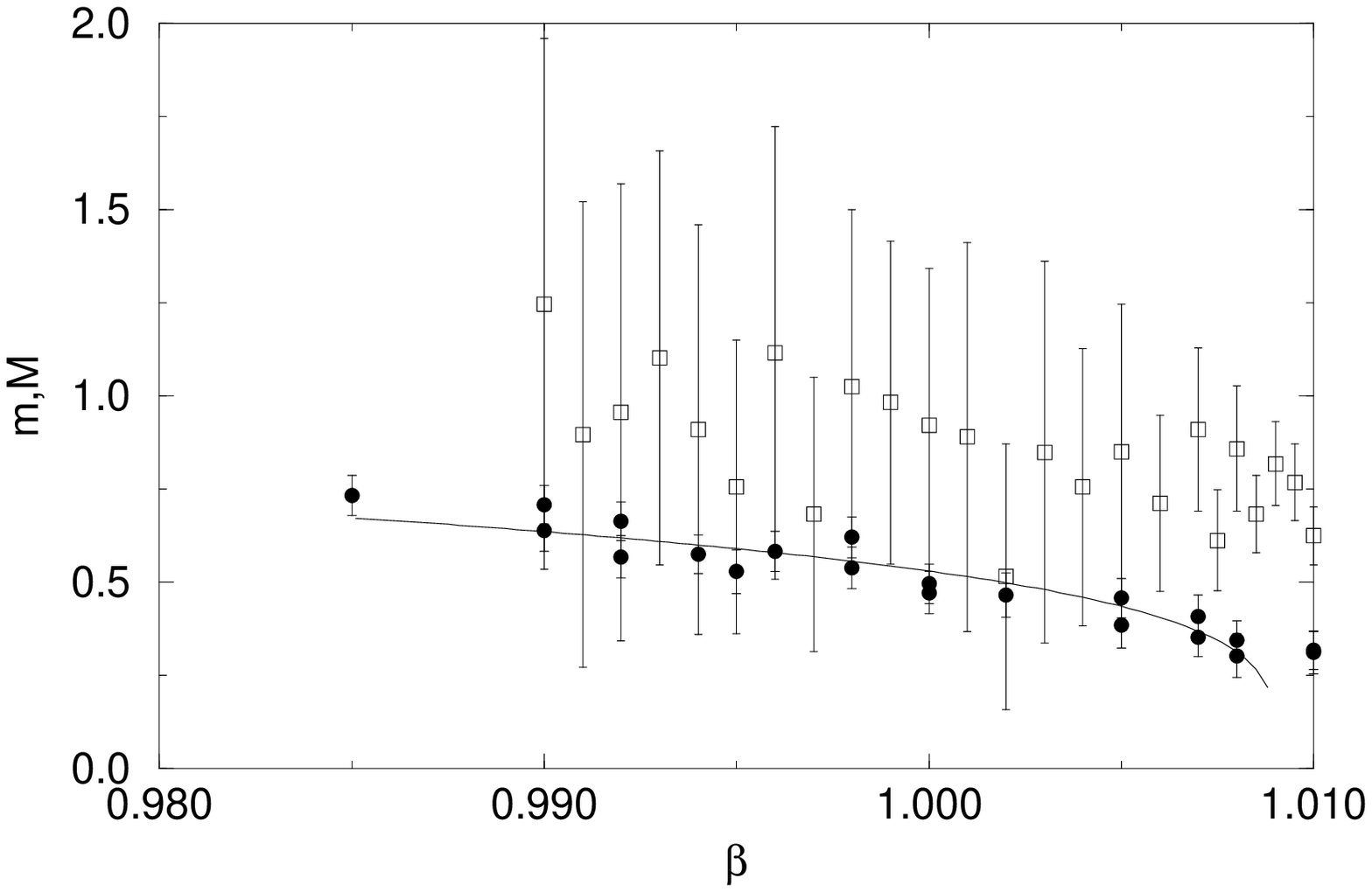}
}
}}
\vskip0.1in\noindent  {\centerline{Fig.6\,
$m$ (circles) and $M$ (squares) vs $\beta$.
}}
\end{minipage}
\vskip0.1in\noindent
\section{The $XY$ model in 3d.}
To further check our disorder parameter we have studied by the same technique
used for $U(1)$ the $XY$ model in 3d, which has a second order phase transition
belonging to the same class of universality as the transition to superfluid
$He_4$.

The field variable is an angle $\theta(i)$ associated to each site. The
action is
\begin{equation}
S = \beta\sum_i\sum_\mu \left[1 - \cos(\Delta_\mu
\theta(i))\right]\label{eq:5.1}\end{equation}
and the partition function
\begin{equation}
Z = \int \prod_i \frac{d\theta(i)}{(2\pi)} \exp(-S) \label{eq:5.2}\end{equation}
The model is compact, so that any change $\theta(i)\to \theta(i) + f(i)$ with
arbitrary $f(i)$ 
leaves correlation functions of compact observables invariant.

As $\beta\to \infty$ $S\simeq \frac{\beta}{2}(\Delta_\mu \theta)^2$ and the
model describes free massless particles. 

At $\beta_c\simeq 0.454$ a second order phase transition takes place, and at
$\beta<\beta_c$ vortices condense. Like in $U(1)$ condensation has been
demonstrated in the literature by a sharp change of density of 
vortices\cite{29}. 
We shall
show instead that for $\beta < \beta_c$ the $U(1)$ symmetry related to
conservation of vortices is spontaneously broken, and we will construct a
disorder parameter to detect the change of symmetry.

Let us define 
\begin{equation}
A_\mu = \partial_\mu \theta \label{eq:5.3}\end{equation}
The invariance under $\theta \to \theta+f$ is  a gauge invariance 
since on $A_\mu$
the transformation is
\[ A_\mu \to A_\mu + \partial_\mu f\]
$A_\mu$ is a gradient, 
eq.(\ref{eq:5.3}), and hence
a pure gauge
\begin{equation}
F_{\mu\nu} = \partial_\mu A_\nu - \partial_\nu A_\mu =  0
\label{eq:5.4}\end{equation}
apart from singularities. 

$\theta(x)$ can be written as a parallel
transport 
\begin{equation}
\theta(x) = \exp(i e \int_{\cal C}^x A_\mu\,d x^\mu)
\label{eq:5.5}\end{equation}
and if eq.(\ref{eq:5.4}) holds this definition is independent of the path $C$.

To investigate possible singularities consider the current
\begin{equation}
j_\mu = \varepsilon_{\mu\alpha\beta} \partial^\alpha A^\beta
\label{eq:5.6}\end{equation}
which is dual to the field strength tensor. $j_\mu$ is identically
conserved (Bianchi identity)
\[ \partial^\mu j_\mu = 0\]
The corresponding charge is
\begin{equation}
Q = \int d^2 x \, j_0(\vec x,t) \label{eq:5.7}\end{equation}
or
\begin{equation}
Q = \int d^2 x\, (\vec\nabla\wedge\vec A) = \oint \vec A\,d\vec x = 2\pi n
\label{eq:5.8}\end{equation}
The path on which the line integral of $\vec A$ is computed is a circle at
infinity: the value $2\pi n$ comes from the definition of $A_\mu$, 
eq.(\ref{eq:5.3}).

In the absence of singularities $j_\mu \equiv 0$ and $Q=0$. 
There exist, however,
configurations with $Q\neq 0$, which are vortices. An example is
\begin{equation}
\tilde \theta_q(\vec x - \vec y) = q\,{\rm atan}\frac{(x-y)_2}{(x-y)_1}
\label{eq:5.9}\end{equation}
For these configurations 
\[A_0=0\qquad \vec A = \frac{q}{|\vec x-\vec y|}\vec \nu_\theta\]
where $\vec \nu_\theta$ is the unit vector tangent to the circle 
$|\vec x-\vec y|=r$. If the field $\vec A$ is the field of velocities the
configuration is a vortex with
\begin{equation}
Q = \oint \vec A\,d\vec x = 2\pi q \label{eq:5.10}\end{equation}
Topology is non trivial.

As a disorder parameter we will use the v.e.v. of the operator which creates a
vortex, following the general rule eq.(\ref{eq:4.2}). Since the conjugate momentum
to $\theta$ as  given by the action is $\Pi = \beta\sin \partial_0\theta$ the
naive definition of $\mu$ would be
\begin{equation}
\mu(\vec y,t) =
\exp\left[-\beta\sum_{\vec n}\sin(\Delta_0\theta(\vec n,t)) \tilde\theta_q(\vec
n-\vec y)\right]
\label{eq:5.11}\end{equation}
The compact version is
\begin{equation}
\mu(\vec y,t) = 
\exp\left[\bar S(\Delta_0\theta(t)-\tilde \theta_q) - \bar S(\Delta_0\theta(t))
\right] \label{eq:5.12}\end{equation}
with $\bar S$ the integral of the lagrangean on time slice $t$
\[ \bar S(t) = \sum_{\vec n} S(\vec n,t)\]
Again, as for the $U(1)$, we compute the correlator
\begin{equation}
{\cal D}(x_0) =
\langle \bar \mu(\vec 0,x_0)\,\mu(\vec 0,0)\rangle 
\mathop\simeq_{|x_0|\to\infty} A\exp(- M |x_0|) + \langle \mu\rangle^2 
\label{eq:5.13}\end{equation}
$\langle\mu\rangle$ is the disorder parameter which signals condensation of
vortices, i.e. spontaneous breaking of the $U(1)$ symmetry. We have from the
definition (\ref{eq:5.12})
\begin{equation}
{\cal D}(x_0) = \frac{Z[S + \Delta S]}{Z[S]} \label{eq:5.14}\end{equation}
where $S+\Delta S$ is obtained from $S$  by the replacements at time slices $0$
and $x_0$ respectively
\begin{eqnarray}
\Delta_0\theta(\vec n,0) &\to& 
\Delta_0\theta(\vec n,0) - \tilde\theta_q(\vec n - 
\vec y)\label{eq:5.15a}\\
\Delta_0\theta(\vec n,0) &\to& 
\Delta_0\theta(\vec n,x_0) + \tilde\theta_q(\vec n - 
\vec y)\label{eq:5.15b}\end{eqnarray}
The change (\ref{eq:5.15a}) can be reabsorbed by a change of variables
\begin{equation}
\theta(\vec n,1) \to \theta(\vec n,1) + \tilde\theta_q
\end{equation}
which leaves the measure unchanged. However it changes 
the space derivatives
at time 1
\begin{equation}
\Delta_i \theta(\vec n,1) \to
\Delta_i \theta(\vec n,1) + \Delta_i\tilde \theta_q\end{equation}
or
\begin{equation}
\vec A(\vec n,1) \to \vec A(\vec n,1) + \Delta_i\tilde \theta_q
\end{equation}
thus adding a vortex to the configuration at $n_0=1$.

The other change in the action coming from the change of variables is
\begin{equation}
\Delta_0\theta(\vec n,1) \to
\Delta_0\theta(\vec n,1) - \tilde\theta_q(\vec n - 
\vec y)\end{equation}
which is like eq.(\ref{eq:5.15a}) with time $0$ replaced by time 1. 
The change of variables can be repeated, producing a vortex at $n_0=2$ and a
shift of $\Delta_0\theta(\vec n,2)$ and so on till $n_0=x_0-1$ when the shift
$-\tilde \theta_q$ of $\Delta S$ disappears with $+\tilde \theta_q$ of 
eq.(\ref{eq:5.15b}).

Again it is numerically convenient to study 
\begin{equation}
\rho(x_0) =\frac{d}{d\beta}{\cal D}(x_0)
\mathop\simeq_{|x_0|\to\infty}
\rho + C e^{-M|x_0|} \end{equation}
with
\begin{equation}
\rho = 2\frac{d}{d\beta}\ln\langle\mu\rangle\qquad
\langle\mu\rangle = \exp\left(
\frac{1}{2}\int_0^\beta\rho(\beta')\,d\beta'\right)
\end{equation}
$\rho$ as a function of $\beta$ is shown in fig.7
for different lattice sizes.
\vskip0.1in\par\noindent
\begin{minipage}{0.99\textwidth}
\epsfxsize = 0.75\textwidth
{\centerline{
\rotatebox{-90}{
{\epsfbox{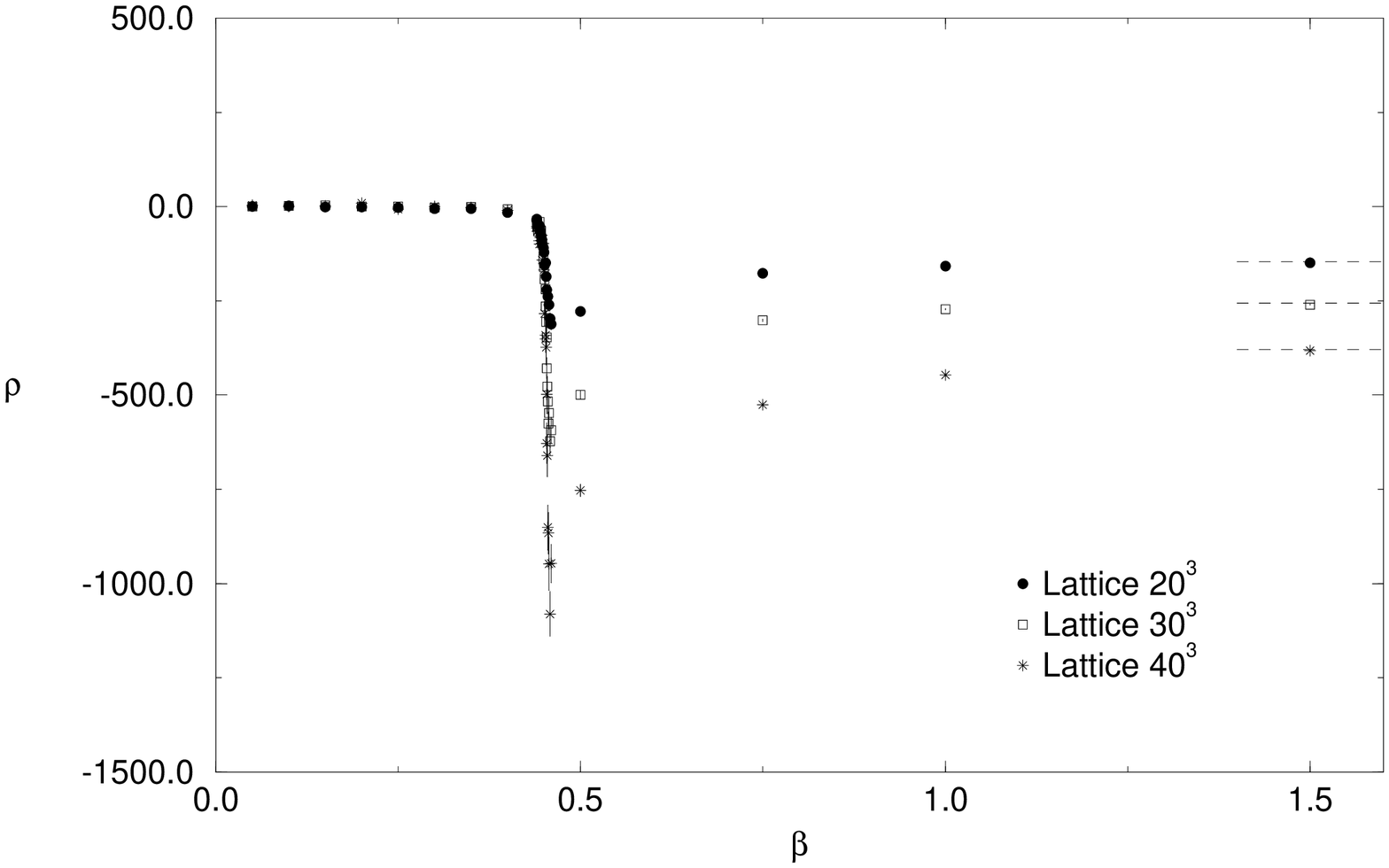}}
}
}}
\vskip0.1in\noindent  Fig.7\,
$\rho$ vs $\beta$ for different lattices. The peak signals the
phase transition.
The lines\\
\phantom{Fig.7}at high $\beta$ are the comparison to perturbation theory, 
eq.(\ref{eq:5.18}).
\end{minipage}
\vskip0.15in\noindent
A huge negative peak signals the phase transition. For $\beta > \beta_c$ the
system describes free particles, the Feynman integral is gaussian and $\rho$
can be computed giving
\begin{equation}
\rho = -11.33 L + 72.7 \label{eq:5.18}\end{equation}
as $L\to\infty$, 
$\rho\to-\infty$, or $\langle\mu\rangle \to 0$, as expected in
the ordered phase in the thermodynamical limit. The agreement with the
prediction (\ref{eq:5.18}) 
at large $\beta$
is shown in the figure: the value (\ref{eq:5.18}) is
represented by the dotted lines.

For $\beta<\beta_c$ $\rho$ tends, as $L\to\infty$, to a finite value compatible
with zero, so that $\langle\mu\rangle\neq 0$: vortices do condense. A finite
size scaling analysis around $\beta_c$ gives the scaling law
\begin{equation}
\frac{\rho}{L^{1/\nu}} = f(L^{1/\nu}(\beta_c-\beta))
\label{eq:5.20}\end{equation}
$\nu$ and $\beta_c$ can be adjusted to satisfy it. The quality of scaling is
shown in fig.8
\par\noindent
\begin{minipage}{0.99\textwidth}
\epsfxsize = 0.70\textwidth
\rotatebox{-90}{
{\centerline{
{\epsfbox{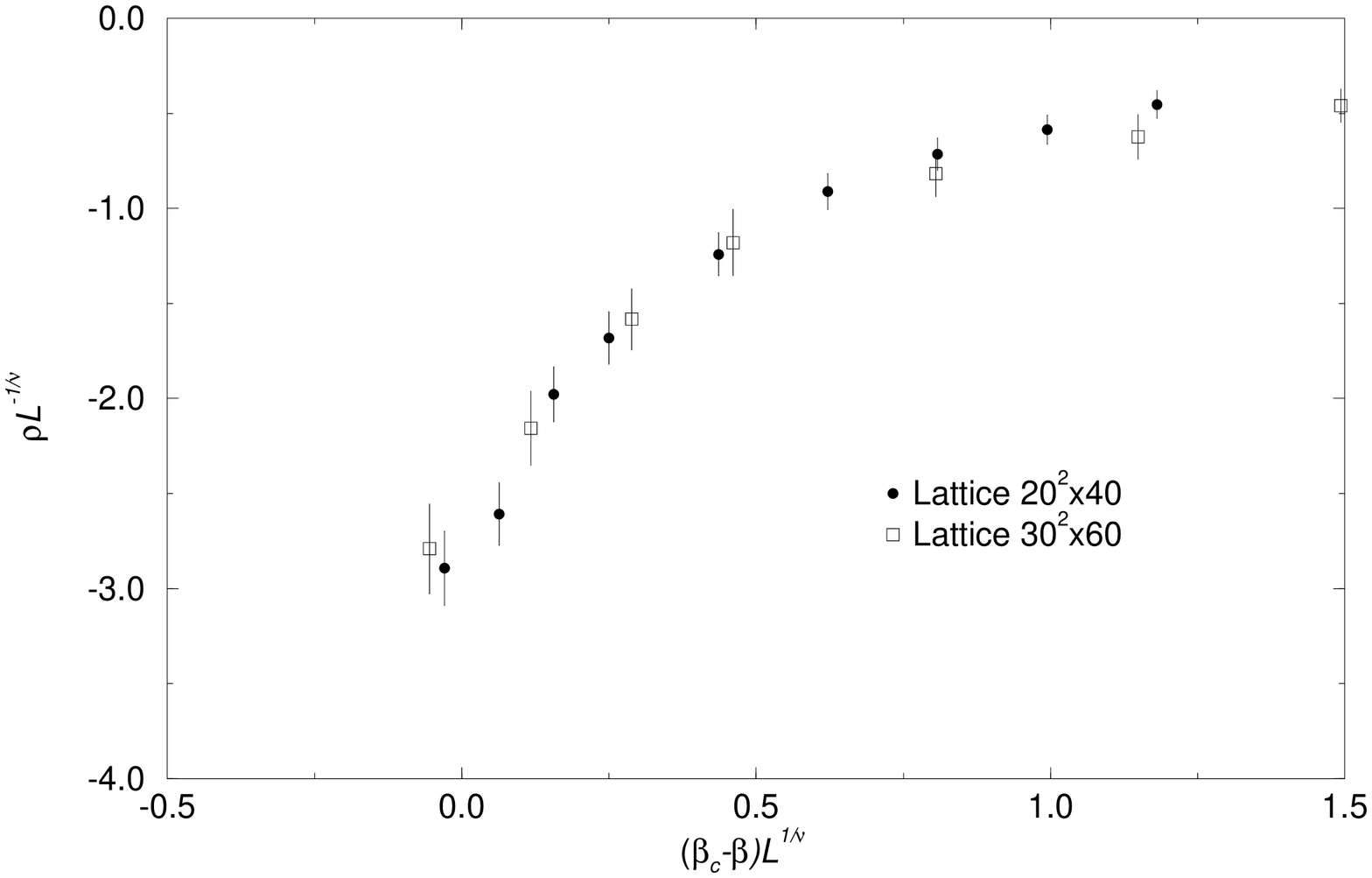}}
}
}}
\vskip-0.3in
\noindent {\centerline{ Fig.8\,
Finite size scaling for $\rho$ in $3d$ $XY$ model.
}}
\end{minipage}
\vskip0.15in\noindent
We get
\begin{eqnarray}
\nu = 0.669\pm 0.065 && [0.670(7)]\label{eq:5.21}\\
\beta_c = 0.4538\pm 0.0003 && [0.4542(2)] \label{eq:5.21b}
\end{eqnarray}
The values in square parantheses have been determined by completely different
methods in the literature\cite{28}, 
and are the critical indices of the transition to
superfluid.  The index $\delta$ of the order parameter
\begin{equation}
\langle\mu\rangle \mathop\simeq_{\beta\sim\beta_c} (\beta_c-\beta)^\delta
\label{eq:5.22}\end{equation}
is also determined
\begin{equation}
\delta = 0.740 \pm 0.029
\end{equation}
Remarkable is the similarity of the shape of $\rho$ in
fig.7 and fig.4 in spite of the fact that the systems have 
nothing to do with each other.

Again we have checked that the construction works, and we will use it to study
the deconfining phase transition as  a possible transition from dual
superconductor to normal.
\section{Monopole condensation vs confinement in QCD.}
Contrary to the Georgi-Glashow model QCD has no fundamental Higgs field.
However, as shown in sect.3 any local operator $\Phi(x)$ in the adjoint
representation has monopoles associated with it, which are located at the zeros
of $\Phi(x)$ for $SU(2)$. For gauge groups of higher rank, say $SU(3)$,
$\Phi(x)$ is written as  a matrix in the fundamental representation
\begin{equation}
\Phi(x) = \sum_a \Phi^a\lambda^a \label{eq:6.1}\end{equation}
with $\lambda^a$ the generators in that representation, monopoles will be
located at the sites where two eigenvalues of the matrix $\Phi(x)$ coincide,
and will be identified by integer charges, in one to one correspondence with
diagonal matrices of the algebra with integer or zero matrix elements\cite{7,8}.
For $SU(N)$ this means $N-1$ $U(1)$ conserved monopole
charges which can condense in the vacuum. For $SU(2)$ there is one charge.
The treatment of higher groups does not add any conceptual point, but only
formal complications. We will therefore use $SU(2)$ formulae to present our
arguments.

The role of $\Phi(x)$, the operator which identifies 
monopoles, can be played  a
priori by infinitely many composite operators of the theory: actually by a
functional infinity of them.
Each of them defines monopoles, which can in principle condense and produce
dual superconductivity. It is not understood a priori to our best knowledge,
if all of these monopole species are really independent of each other and 
 if many
of them could condense at the same time in connection 
with confinement. It could also
be that  $U(1)$ superconductivity for many monopole species
is a manifestation of a more clever mechanism, a really non abelian
superconductivity.
This would implement 
the guess by t'Hooft, that all monopole species, defined by any operator
$\Phi(x)$ are physically equivalent\cite{9}.

What we can presently do is to investigate these issues on the lattice. 
Looking at the problem from the point of view of symmetry is the most
direct way.

The choices which have been suggested in the literature for $\Phi(x)$ 
are\cite{9}
\begin{itemize}
\item[1)] The Polyakov line, i.e. the parallel transport along the time axis to
$+\infty$ and back from $-\infty$ via periodic boundary conditions.
\item[2)] Any component of the field strength.
\item[3)] $F_{\mu\nu}F_{\mu\nu}$, at least for $SU(3)$ when it has an octet
part. For $SU(2)$ it is a singlet.
\item[4)] The operator which is implicitely defined by the maximization of the
quantity, with respect to gauge variations\cite{9,30}
\begin{equation}
{\rm Tr}\left\{ \sigma_3 \Omega U_\mu(n) \Omega^\dagger\sigma_3
\Omega^\dagger U^\dagger_\mu(n) \Omega\right\} = {\rm max}
\label{eq:6.6}\end{equation}
with $\Omega(n)$ a generic gauge transformation. This procedure defines an
operator $\Phi$ in the adjoint representation which coincides with
\begin{equation}
\sum_\mu\left( U^\dagger(n) \sigma_3 U_\mu(n) +
U_\mu(n-\hat\mu) \sigma_3 U^\dagger_\mu(n-\mu)\right) \end{equation}
in the maximal abelian gauge.
The explicit form in a generic representation is not known.
\end{itemize}
To explore dual superconductivity we will construct a disorder parameter, as
the v.e.v. $\langle\mu\rangle$ of an operator $\mu$ with $U(1)$ magnetic
charge, as we have done for $U(1)$ and $XY$ model.

Whatever the choice of $\vec \Phi$, in the gauge in which
$\vec\Phi\cdot\vec\sigma$ is diagonal, i.e. after abelian projection, any link
$U_\mu(n)$ can be written\cite{30}
\begin{eqnarray}
U_\mu(n) &=& e^{i \sigma_3 \alpha_\mu(n)}
e^{i \sigma_2 \gamma_\mu(n)}e^{i \sigma_3 \beta_\mu(n)}\nonumber\\
&&\left[ e^{i \sigma_3 \alpha_\mu(n)}
e^{i \sigma_2 \gamma_\mu(n)}e^{-i \sigma_3 \alpha_\mu(n)}\right]
e^{i \sigma_3 (\alpha_\mu(n)+\beta_\mu(n))}
\end{eqnarray}
$e^{i \sigma_3 \theta_\mu(n)}\equiv
e^{i \sigma_3 (\alpha_\mu(n)+\beta_\mu(n))}$ is the abelian link, which
parallel transports the $U(1)$ field related to the monopole 
charges defined by
$\hat\Phi$.

The creation operator of a monopole at time $t=0$ will be defined by changing
the kinetic term of the action at that time by 
adding the field of the monopole
to $\theta^{i0}(n)$\cite{30a}.

The change on the $i-0$ plaquette $\Pi_{i0}$
will be
\[ \Pi_{i0}(\vec n,0) \to \Pi'_{i0}(\vec n,0)_b\]
\begin{eqnarray*}
\Pi_{i0}(\vec n,0) &=&
{\rm Tr}\left[
U_i(\vec n,0) U_0(\vec n + \hat i),0) U^\dagger_i(\vec n,1) U^\dagger_0(\vec n,0)
\right]\\
{\Pi'}_{i0}(\vec n,0)_b&=&
{\rm Tr}\left[
U'_i(\vec n,0) U_0(\vec n + \hat i),0) U^\dagger_i(\vec n,1) U^\dagger_0(\vec n,0)
\right]
\end{eqnarray*}
and
\begin{equation}
U'_i(\vec n,0) =
e^{i \Lambda(n)\sigma_3} U_i(\vec n,0) e^{i b_i^\perp\sigma_3}
e^{-i \Lambda(n)\sigma_3} \label{eq:6.7}\end{equation}          
We have operated the separation of the vector potential describing the
monopole into transverse and longitudinal part
\[ b_i(n) = b_i^\perp(n) + \partial_i\Lambda(n) \qquad
\partial_i b_i^\perp(n) = 0\]
The gauge part $e^{i\Lambda(n)\sigma_3}$ and $e^{i\Lambda(n+1)\sigma_3}$
can be reabsorbed by a rotation of the $U_0$'s which leaves the functional
measure invariant, so that the definition is independent on the choice of the
classical gauge for $b_i(n)$, and the net effect    is to add $b_i^\perp(n)$
to the abelian phase of $U_i(\vec n,0)$.

Consider now the correlator
\begin{equation}
{\cal D}(t) =
\langle \bar\mu(\vec x, t) \mu(\vec x,0)\rangle\label{eq:6.8}\end{equation}
As usual we can write
\begin{equation}
{\cal D}(t) = \frac{Z[S + \Delta S]}{Z[S]} \label{eq:6.9}\end{equation}
where $S+\Delta S$ is obtained by the substitution
\begin{eqnarray*}
\Pi^{0i}(\vec n,0) &\to & \Pi^{0i}_b(\vec n,0)\\
\Pi^{0i}(\vec n,x_0) &\to & \Pi^{0i}_{-b}(\vec n,x_0)
\end{eqnarray*}
The subscript $b$, $-b$ is to recall the sign of the monopole charge.

The effect of this procedure is to have a monopole created at $t=0$, which
propagates to time $t$, when it disappears. The construction 
to show this
is identical to
$U(1)$ if $\Phi$ is the Polyakov line. There after abelian projection the
temporal links are diagonal 
\[ U_0(n) = e^{i \sigma_3 \alpha_0(n)}\]
and
the operator $e^{i \sigma_3 b^\perp_i}$ in the definition of $U'_i$ commutes
with them.

Then a change of variables from $U_i$ to $U'_i$ 
which leaves the measure invariant
brings 
$\Pi_{i0}$ back to the original form, but changes $\Pi_{ij}(\vec n,1)$ by
adding a monopole to the abelian field
\[\Delta_i\theta_j - \Delta_j\theta_i
\to \Delta_i\theta_j - \Delta_j\theta_i +
\Delta_i b_j - \Delta_j b_i\]
The other change is to $\Pi^{i0}(\vec n,1)\to \Pi^{i0}_b(\vec n,1)$.
The procedure can then be repeated till at time $t$ the change is reabsorbed by
$-b$.

If the $U_0$'s are 
not diagonal in the abelian projected gauge, the construction is
the same modulo additional parallel transports
in the change of variables, which  do not modify its
content.

Again one can either measure ${\cal D}$ itself or
\begin{equation}
\rho(t) = \frac{d}{d\beta}
\ln {\cal D}(t) = \langle S\rangle_S -
\langle S+\Delta S\rangle_{S+\Delta S}\end{equation}
One expects
\begin{eqnarray}
{\cal D}(t) &=& \langle\mu\rangle^2 + A e^{-M t}\label{eq:6.10a}\\
\rho(t) &\simeq& \rho + C e^{-M t}\label{eq:6.10b}\\
\rho &=& 2 \frac{d}{d\beta}\ln\langle\mu\rangle\label{eq:6.10c}
\end{eqnarray}
A similar analysis to the one performed for $U(1)$ brings to the determination
of $\langle\mu\rangle$ in the thermodinamical limit, and of $M$.
$\langle\mu\rangle\neq 0$ signals dual superconductivity. A measurement of the
penetration depth of the field allows to establish the type of superconductor.

At finite temperature, i.e. keeping the time extension $N_T$ of the lattice
much smaller than the space extension $N_S$, the correlation of
$\langle\mu\rangle$ to confinement can be studied.
There a single monopole is used, and $\langle\mu\rangle$ 
is measured. A careful
examination of the construction given above shows that, when computing 
$Z[S+\Delta S]$ of eq.(\ref{eq:6.8})
if periodic boundary conditions are used, the change of variable 
described above
adds more
and more monopoles when we go trough the boundary in time.
The way to have one monopole is to use antiperiodic 
boundary conditions in time.
This suggests that monopoles behave as fermions.

We are systematically exploring by our disorder parameter the deconfining
transition in $SU(2)$ and $SU(3)$, by different choices for $\vec \Phi$, the
field which defines monopoles. 
We are measuring penetration depths and critical
indices. 
As for $\Phi$ we consider Polyakov line, field strength component
and max abelian projection.
Typical behaviour for $\rho$ are shown in fig.9 and fig.10
for $SU(2)$ and $SU(3)$
\vskip0.1in\noindent
\begin{minipage}{0.99\textwidth}
\epsfxsize = 0.80\textwidth
{\centerline{
{\epsfbox{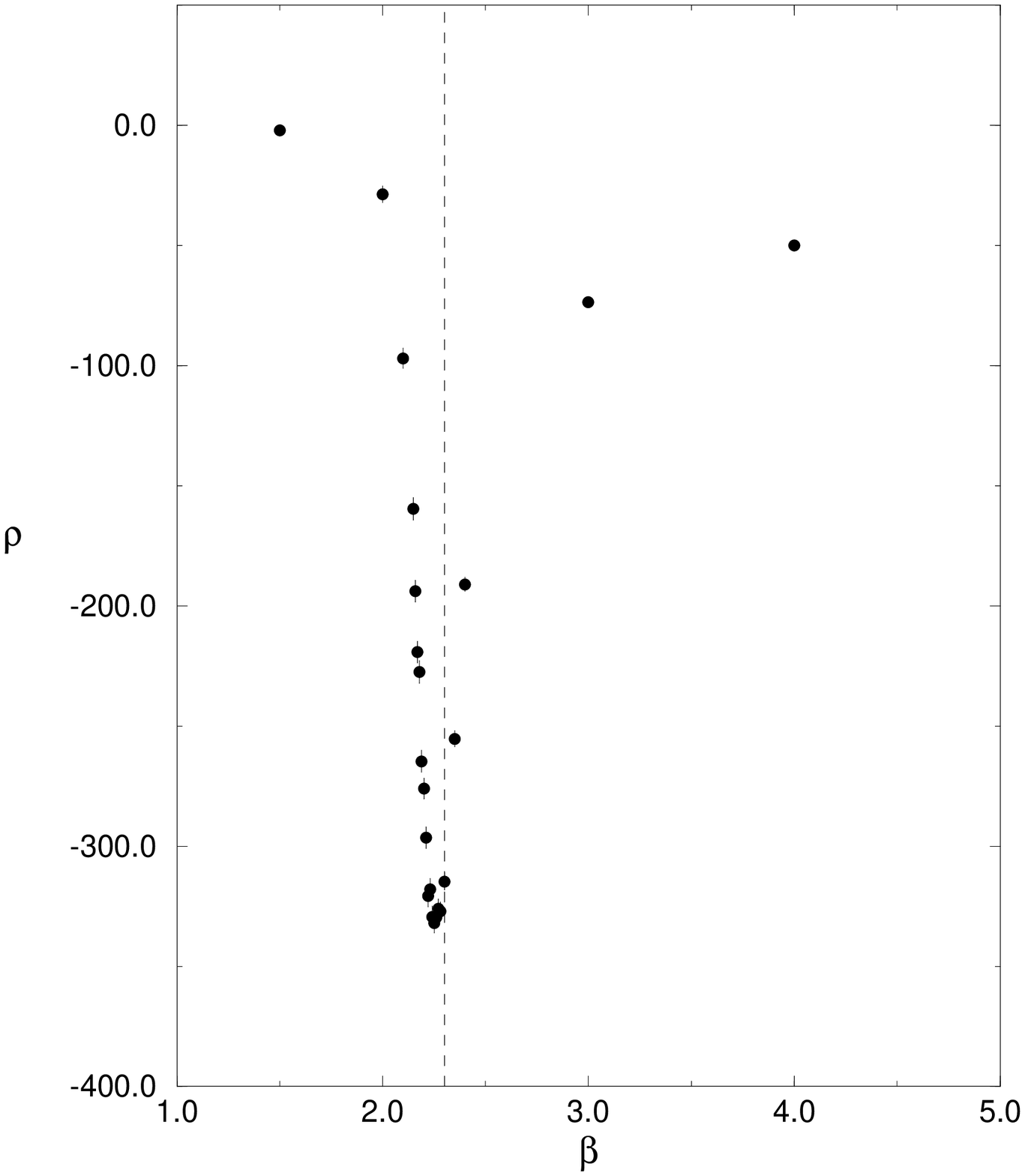}
}
}}
\vskip0.1in\noindent  Fig.9\,
$\rho$ vs $\beta$ for $SU(3)$ gauge theory. The peak signals
deconfining phase transition.\\
\phantom{Fig.9}Here monopoles
are defined by the abelian projection on Polyakov line.
\end{minipage}
\vskip0.2in\noindent
Preliminary evidence is that for all the species of monopoles considered,
vacuum behaves as a dual superconductor, and undergoes a phase transition to
normal at the deconfinement point.
This supports the guess of t'Hooft about physical equivalence of different
monopole species\cite{9}.

\begin{minipage}{0.99\textwidth}
\epsfxsize = 0.99\textwidth
{\centerline{
{\epsfbox{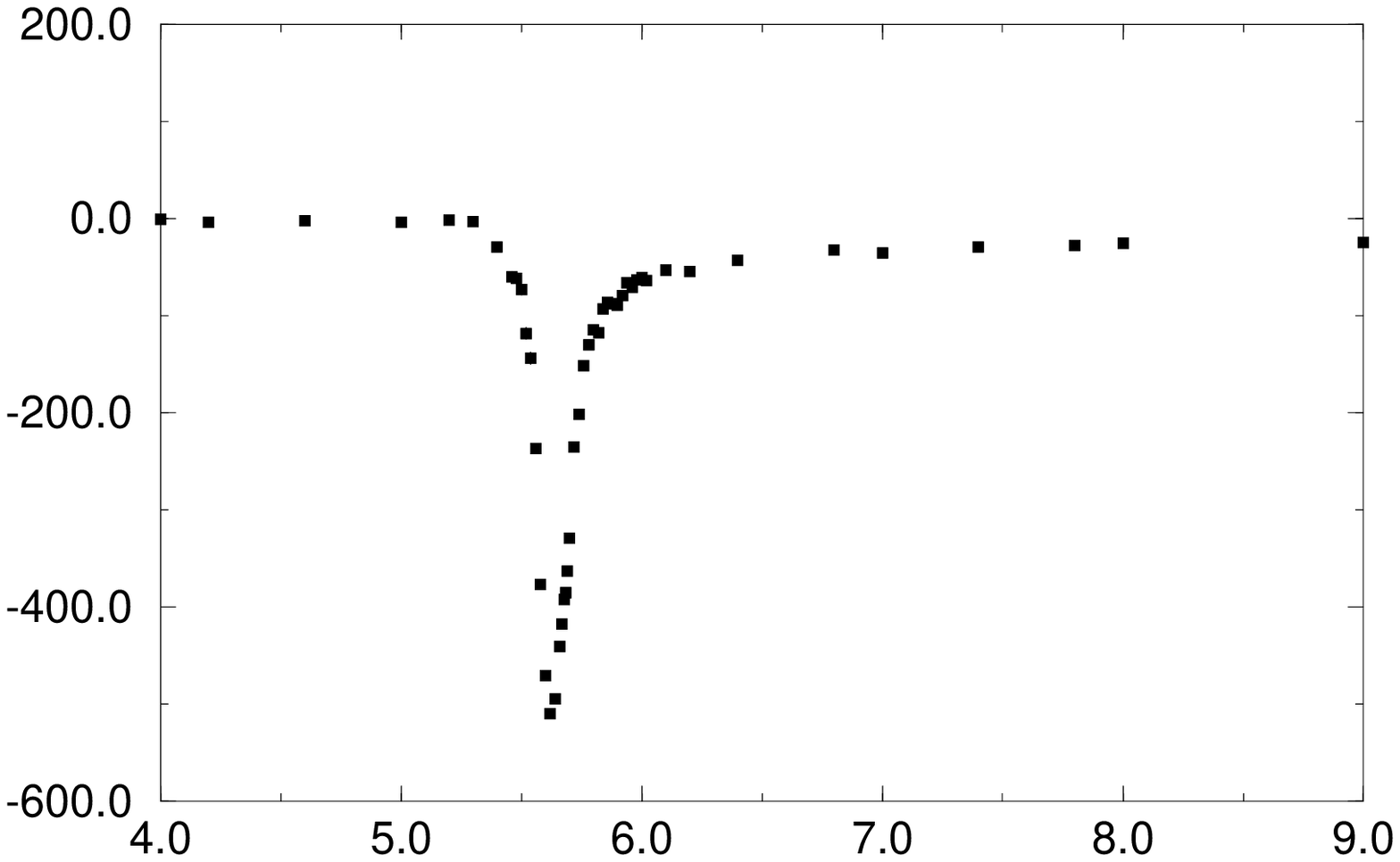}
}
}}
\vskip0.1in\noindent {\centerline{ Fig.10\,
Same as Fig.9 for $SU(2)$.
}}
\end{minipage}
\vskip0.15in\noindent

\section{Concluding remarks.}
Our strategy to answer the question if dual superconductivity is the mechanism
of colour confinement, is to look at the symmetry of the vacuum. For that we
have constructed a disorder parameter, which directly detects dual
superconductivity.

The construction has been tested in known systems, like the $U(1)$ compact
gauge theory and  the $XY$ model in 3 dimensions.

We have evidence that for many choices of the 
effective field $\Phi(x)$ defining
monopole species, 
dual superconductivity is present in the confined phase, and
disappears in the quark gluon phase.

Additional relevant information from lattice is that
\begin{itemize}
\item[1)] Flux tubes exist in the space between propagating  $Q\bar Q$ 
pair\cite{31,35}.
\item[2)]
If one single species of monopoles were at work to produce superconductivity,
then the electric field 
in the Abrikosov tubes should be the field
of the $U(1)$ group to which monopole charges belong. An  analysis of the
colour content of the flux tubes shows instead that its direction in colour
space is uncorrelated to the direction of $\vec \Phi$\cite{35,34}.
\end{itemize}
Moreover, whatever the abelian projection is, there exist one gluon in $SU(2)$,
two of them in $SU(3)$, which have zero electric charge with respect to the
residual $U(1)$'s, and therefore cannot be confined. The adjoint string tension
is zero, and also this fact seems to contradict lattice observations\cite{35}.

A different approach to the problem is to pay less attention to symmetry, and
look at more quantitative facts, like abelian dominance\cite{10} 
and monopole dominance\cite{11}.

The abelian part of the field as defined by abelian projection, in the max
abelian gauge, is a good approximation to full dynamics, 
and of it the contribution
of monopoles is dominant.
In a sense abelian dominance is expected, since, after maximal 
abelian projection
the links are diagonal within 85\%. However the fact that it happens is surely
relevant. 

Looking at symmetry, as we do, is a complementary approach. 

Hopefully
a picture will emerge from all these efforts, which will improve our
understanding of the theory.


\begin{thebibliography}{99}
\bibitem{1} See eg. A.H.~M\"uller, \NP{B250,1985,327.}
\bibitem{2}G. 't~Hooft, in ``High Energy Physics'', EPS
International Conference, Palermo 1975, ed. A.~Zichichi.
\bibitem{3}S. Mandelstam, {Phys. Rep.} {\bf 23C} (1976), 245.
\bibitem{4}G. Parisi, { Phys. Rev.} {\bf D11}  (1975), 971.
\bibitem{5} S. Weinberg, {Progr. of Theor. Phys.
Suppl.\/} {\bf 86} (1986), 43.
\bibitem{6}N.N.~Bogolubov, \JL{N.C.,7,1958,794}.
\bibitem{6a}J.G.~Valatin, \JL{N.C.,7,1958,843}.
\bibitem{7}C. Goddard, J. Nuyts, P. Olive,
\NP{B125, 1977,1}.
\bibitem{8}S. Coleman, {Erice Lectures 1981}, Plenum Press,
Ed. A. Zichichi.
\bibitem{9} G. 't~Hooft, \NP{B190,1981, 455}.
\bibitem{10}T. Suzuki, {Nucl. Phys.}(Proc. Suppl.) {\bf B30} (1993), 176.
\bibitem{11} R.J. Wensley, J.D. Stack \PRL{63,1989, 1764}.
\bibitem{12}A.B. Abrikosov, JETP {\bf 5}  (1957), 1174.
\bibitem{13}P.A.M. Dirac: {Proc. Roy. Soc.} (London), Ser. A, {\bf 133}
(1931), 60.
\bibitem{14} L.P. Kadanoff and H. Ceva, \PR{B3,1971,3918}.
\bibitem{15} G. 't~Hooft, \NP{B79,1974,276}.
\bibitem{16}A.M. Polyakov, {JEPT Lett.} {\bf 20} (1974), 894.
\bibitem{17}A. Di Giacomo and M. Mathur, \PL{B400,1997,129}.
\bibitem{a17}A. Di Giacomo and G.Paffuti, \PR{D56,1997,6816}.
\bibitem{18} E.C. Marino, B. Schror and J.A. Swieca, 
\NP{B200,1982,473}.
\bibitem{18a}L.Del Debbio, A.Di Giacomo and G.Paffuti, 
\PL{B 349,1995,513}.
\bibitem{18b}L.Del Debbio, A.Di Giacomo, G.Paffuti and P.Pieri, 
\PL{B355,1995,255}.
\bibitem{20} J. Fr\"ohlich and P.A. Marchetti, 
\CMP{112,1987,343}.
\bibitem{b17}A. Liguori, M.~Mintchev and M. Rossi
\PL{B305,1993,52}.
\bibitem{22}T. De Grand, D. Toussaint,
\PR{D22,1980,2478}.
\bibitem{23}K.G. Wilson: \PR{D10,1974, 2445 }.
\bibitem{24}J.Jersak, O.Lang and J. Neuhaus,
\PRL{77,1996,1933}.
\bibitem{24a}W. Kerler, C. Rebbi and A. Weber
\NP{B 450,1995 670}.
\bibitem{26}V. Singh, R.W. Haymaker, D.A. Brown, 
\PR{D47,1993,1715}.
\bibitem{25} V. Cirigliano, G.Paffuti,{\em Magnetic Monopoles in U(1) lattice gauge theory} Hep-Lat 9707219.
\bibitem{27}G. Di Cecio, A. Di Giacomo, G. Paffuti and M. Trigiante,
\NP{B489,1997,739}.
\bibitem{29}G. Kohning, R.E. Schrook and P.Willis,
\PRL{57,1986,1358}.
\bibitem{28}A.P. Gottlob, M. Hasenbush, CERN-TH 6885-93.
\bibitem{30}A.S. Kronfeld, G. Schierholz and U.J. Wiese
\NP{B293,1987,461}.
\bibitem{30a}
A.Di Giacomo, B. Lucini, L. Montesi and G.Paffuti,
hep-lat 9709005, to appear in the proceedings of LATTICE97.
\bibitem{31} R.W. Haymaker, J. Wosiek, 
\PR{D36,1987,3297}.
\bibitem{35} A. Di Giacomo, M.~Maggiore and \v{S}.~Olejn\'{\i}k,
\PL{B236,1990,199};\NP{B347,1990,441}.
\bibitem{34}J. Greensite and J. Winchester, 
\PR{D40,1989,4167}.
\bibitem{36}J. Ambjorn, P. Olesen and C.Peterson,
\NP{B240,1984,189}.
\end{thebibliography}
\end{document}